\documentclass[longauth]{aa}  
\usepackage{graphicx}
\usepackage{txfonts}
\usepackage{array}
\begin{document}
   \title{The VIMOS VLT Deep Survey}

   \subtitle{Evolution of the major merger rate since $z \sim 1$ from
             spectroscopically confirmed galaxy pairs.
\thanks{based on observations
         obtained with the European Southern Observatory Telescopes at the
         Paranal Observatory, under programs 072.A-0586 and 073.A-0647}
}

   \author{
L. de Ravel\inst{1} 
\and O. Le F\`evre\inst{1}
\and L. Tresse \inst{1}
\and D. Bottini \inst{2}
\and B. Garilli \inst{2}
\and V. Le Brun \inst{1}
\and D. Maccagni \inst{2}
\and R. Scaramella \inst{4,13}
\and M. Scodeggio \inst{2}
\and G. Vettolani \inst{4}
\and A. Zanichelli \inst{4}
\and C. Adami \inst{1}
\and S. Arnouts \inst{23,1}
\and S. Bardelli  \inst{3}
\and M. Bolzonella  \inst{3} 
\and A. Cappi    \inst{3}
\and S. Charlot \inst{8,10}
\and P. Ciliegi    \inst{3}  
\and T. Contini \inst{7}
\and S. Foucaud \inst{21}
\and P. Franzetti \inst{2}
\and I. Gavignaud \inst{12}
\and L. Guzzo \inst{9}
\and O. Ilbert \inst{20}
\and A. Iovino \inst{9}
\and F. Lamareille \inst{7}
\and H.J. McCracken \inst{10,11}
\and B. Marano     \inst{6}  
\and C. Marinoni \inst{18}
\and A. Mazure \inst{1}
\and B. Meneux \inst{22,24}
\and R. Merighi   \inst{3} 
\and S. Paltani \inst{15,16}
\and R. Pell\`o \inst{7}
\and A. Pollo \inst{1,17}
\and L. Pozzetti    \inst{3} 
\and M. Radovich \inst{5}
\and D. Vergani \inst{2}
\and G. Zamorani \inst{3} 
\and E. Zucca    \inst{3}
\and M. Bondi \inst{4}
\and A. Bongiorno \inst{22}
\and J. Brinchmann \inst{19}
\and O. Cucciati \inst{9,14}
\and S. de la Torre \inst{1}
\and L. Gregorini \inst{4}
\and P. Memeo \inst{2}
\and E. Perez-Montero \inst{7}
\and Y. Mellier \inst{10,11}
\and P. Merluzzi \inst{5}
\and S. Temporin \inst{9}
}
   
\offprints{L. de Ravel : loic.deravel@oamp.fr}

\institute{
Laboratoire d'Astrophysique de Marseille, UMR 6110 CNRS-Universit\'e de Provence, BP8, F-13376 Marseille Cedex 12, France 
\and 
IASF-INAF, Via Bassini 15, I-20133, Milano, Italy
\and
INAF-Osservatorio Astronomico di Bologna, Via Ranzani 1, I-40127, Bologna, Italy 
\and 
IRA-INAF, Via Gobetti 101, I-40129, Bologna, Italy 
\and 
INAF-Osservatorio Astronomico di Capodimonte, Via Moiariello 16, I-80131, Napoli, Italy 
\and 
Universit\`a di Bologna, Dipartimento di Astronomia, Via Ranzani 1, I-40127, Bologna, Italy 
\and 
Laboratoire d'Astrophysique de Toulouse-Tarbes, Universit\'e de Toulouse, 
CNRS, 14 av. E. Belin, F-31400 France
\and 
Max Planck Institut f\"ur Astrophysik, D-85741, Garching, Germany 
\and 
INAF-Osservatorio Astronomico di Brera, Via Brera 28, I-20021, Milan, Italy 
\and
Institut d'Astrophysique de Paris, UMR 7095, 98 bis Bvd Arago, F-75014, Paris, France 
\and 
Observatoire de Paris, LERMA, 61 Avenue de l'Observatoire, F-75014, Paris, France 
\and 
Astrophysical Institute Potsdam, An der Sternwarte 16, D-14482, Potsdam, Germany 
\and 
INAF-Osservatorio Astronomico di Roma, Via di Frascati 33, I-00040, Monte Porzio Catone, Italy 
\and 
Universit\'a di Milano-Bicocca, Dipartimento di Fisica, Piazza delle Scienze 3, I-20126, Milano, Italy
\and
Integral Science Data Centre, ch. d'\'Ecogia 16, CH-1290, Versoix, Switzerland 
\and
Geneva Observatory, ch. des Maillettes 51, CH-1290, Sauverny, Switzerland 
\and
The Andrzej Soltan Institute for Nuclear Studies, ul. Hoza 69, 00-681 Warsaw, Poland
\and 
Centre de Physique Th\'eorique, UMR 6207 CNRS-Universit\'e de Provence, F-13288, Marseille, France 
\and 
Centro de Astrof{\'{i}}sica da Universidade do Porto, Rua das Estrelas, P-4150-762, Porto, Portugal 
\and
Institute for Astronomy, 2680 Woodlawn Dr., University of Hawaii, Honolulu, Hawaii, 96822, USA 
\and
School of Physics \& Astronomy, University of Nottingham,
University Park, Nottingham, NG72RD, UK
\and
Max Planck Institut f\"ur Extraterrestrische Physik (MPE), Giessenbachstrasse 1,
D-85748 Garching bei M\"unchen,Germany
\and
Canada France Hawaii Telescope corporation, Mamalahoa Hwy,  
Kamuela, HI-96743, USA
\and
Universit\"atssternwarte M\"unchen, Scheinerstrasse 1, D-81679 M\"unchen, Germany\\
\\
\email{loic.deravel@oamp.fr}
}
   \date{Received 11/07/2008; accepted 31/01/2009}


\abstract
{The rate at which galaxies grow via successive mergers is a key element to understand the main phases of galaxy evolution.}
{We measure the evolution of the fraction of galaxies in pairs and the merging rate since redshift $z \sim 1$ assuming a ($H_0=70 km s^{-1} Mpc^{-1}, \Omega_{M}=0.3$ and $\Omega_{\Lambda}=0.7$) cosmology.}
{From the VIMOS VLT Deep Survey we use a sample of 6464 galaxies with $I_{AB} \leq 24$ to identify 314 pairs of galaxies, each member with a secure spectroscopic redshift, which are close in both projected separation and in velocity.}
{We estimate that at $z\sim 0.9$, $10.9 \pm 3.2\%$ of galaxies with $M_B(z) \leq -18-Qz$ ($Q=1.11$) are in pairs with 
separations $\Delta r_p \leq 20h^{-1}\ kpc$, $\Delta v \leq 500$ km/s, and with $\Delta M_B \leq 1.5$,
significantly larger than $3.8 \pm 1.7 \%$ at $z \sim 0.5$; thus, the pair fraction evolves as $(1+z)^m$ with $m = 4.73 \pm 2.01$. For bright galaxies with $M_B(z=0) \leq -18.77$, the pair fraction is higher and its evolution with redshift is flatter with $m=1.50 \pm 0.76$, a property also observed for galaxies with increasing stellar masses. Early-type pairs (dry mergers) increase their relative fraction from $3\%$ at $z \sim 0.9$ to $12\%$ at $z \sim 0.5$. 
The star formation rate traced by the rest-frame [OII] $EW$ increases by $26 \pm 4\%$ for pairs with the smallest separation $r_p \leq 20h^{-1}kpc$. Following the prescription to derive merger timescales of Kitzbichler \& White (2008) we find that the merger rate of $M_B(z) \leq -18-Qz$ galaxies evolves as $N_{mg}=(4.96 \pm 2.07)\times 10^{-4}) \times (1+z)^{2.20 \pm 0.77}  mergers\ Mpc^{-3} Gyr^{-1}$.}
{The merger rate of galaxies with $M_B(z) \leq -18-Qz$ has significantly evolved since $z \sim 1$ and is strongly dependent on the luminosity or stellar mass of galaxies. The major merger rate increases more rapidly with redshift for galaxies with fainter luminosities or stellar mass, while the evolution of the merger rate for bright or massive galaxies is slower, indicating that the slow evolution reported for the brightest galaxies is not universal. The merger rate is also strongly dependent on the spectral type of galaxies involved. Late-type mergers were more frequent in the past, while early-type mergers are more frequent today, contributing to the rise in the local density of early-type galaxies. About $20\%$ of the stellar mass in present day galaxies with $log(M/M_{{\odot}}) \geq 9.5$ has been accreted through major merging events since $z = 1$. This indicates that major mergers have contributed significantly to the growth in stellar mass density of bright galaxies over the last half of the life of the Universe.} 


\maketitle
%

\section{Introduction}

In the current hierarchical structure formation paradigm, the mass assembly 
in galaxies proceeds via a process of coalescence between increasingly more massive 
dark matter halos. This halo merging tree history can be 
quantified by a  halo merger rate, measuring the growth of mass per average mass in a 
representative volume of the Universe. However, these models do not directly 
predict a growth of galaxy mass via mergers (\cite{moore}), and the actual contribution
of mergers to the evolution of galaxies remains poorly predicted. 

Merging two galaxies is potentially a very powerful process.
It is possible that during major merger events, i.e. mergers where the two components have more or less the same mass, disks could be transformed into spheroidals, as predicted using detailed simulations (\cite{combes}; \cite{mihos}; \cite{conselice06}).
It is also expected that major merger events 
profoundly modify the spectrophotometric
properties of the galaxies involved, for instance triggering a burst of 
star formation (e.g. \cite{P05}).
Galaxies in the process of merging are observed, however the contribution
of this process to the evolution of the global galaxy population is not
yet precisely constrained. 
Indirect evidence for merging is also inferred from other galaxy properties
like the luminosity or mass function. The luminosity of the red
bulge dominated population of galaxies is measured to increase
since $z\sim1$, part of which could be produced by mergers (\cite{ilbert06}).
it seems possible that the increase in the density of intermediate mass early-type 
galaxies since $z\sim1$ may be happening at the expense of late-type galaxies involved in 
merging events (\cite{tresse}).
Merging is therefore potentially a very important physical phenomenon which
could drive the evolution of galaxies along cosmic time. 
The average numbers of merger events needed to build a 
typical $M_*$ galaxy, the contribution of mergers to the mass growth of galaxies, 
or the identification of a prefered time in the life of the
Universe when mergers were more frequent, 
are all important elements to help towards our understanding of galaxy evolution. 
It is then crucial to quantify the contribution of merging to the evolution process
and its impact on  important quantities
like the cosmic star formation rate (e.g. \cite{bouwens}; \cite{tresse}; \cite{woods}) 
or the global stellar mass density (e.g. \cite{Arnouts}; \cite{bundy}; \cite{pozzetti}).

To estimate the contribution of mergers to the formation and evolution of 
galaxies is not a trivial task. In the nearby Universe merger events can be 
identified aposteriori from perturbed morphologies, wisps, tails, and other peculiar 
signatures seen at low surface brightness. Only recently volume complete 
measurements of the merger rate in the nearby Universe are becoming available. 
In the Millennium catalogue, especially tailored to a volume complete identification
of merging events, de Propris et al. (2007) use the relative velocity measured from spectroscopic redshifts to 
confirm true galaxy pairs in the process of merging. They find 
that the merger fraction is 2\% at a mean redshift of 0.06, refining
earlier estimates based on pair fraction (\cite{P00}; \cite{P02}). 
At higher redshifts,
searching for evidence for past mergers becomes increasingly difficult,
because the residual signatures of mergers often have a too low surface brightness. 
At redshifts $z\geq0.3$, it is therefore easier to search for 'apriori mergers',
encounters that are likely to lead to a merger event, rather than to look for 
'aposteriori' signs of past mergers. When two galaxies are close together in
space, and depending on their relative velocities, gravity is acting to bring them
closer for a bound system that will merge. A measure of the merging frequency is then
to count galaxy pairs with a separation and velocity difference such as they 
are likely to be gravitationally bound and destined to merge. By selecting pairs of galaxies  
with similar magnitudes and hence approximately with similar masses, one can focus on
major merger events. They are able to significantly contribute to the mass assembly, 
to modify morphologies, as well as to significantly alter the star and gas
content of the incoming galaxies. Assuming that a 
dynamically bound system of two galaxies will most likely evolve into one more massive 
galaxy, one can then derive the merger rate from the pair count. A major 
uncertainty of this estimator is the timescale upon which a merger will be 
completed. N-body simulations are then used to provide reasonable estimates
of the merger timescales (\cite{conselice06}; Kitzbichler \& White, 2008). 

Cold dark matter simulations show that the evolution of the dark matter halos 
merger rate follows a power law  $N_{mg} = N_{mg,0}(1+z)^m$ 
where $N_{mg,0} = N_{mg}(z=0)$ is the local value, and $m$ parameterizes the evolution.
While some simulations predict that $m$ should have $2.5 \leq m \leq 3.5$ (Gottl\"\o ber, 2001),
measuring $m$ directly from galaxy samples
is an important step to understand the evolution of galaxies.
Many observational attempts have been carried out to track the evolution of the merger 
rate as a function of redshift (e.g. \cite{burkey}; \cite{carlberg94}; \cite{yee}; \cite{P97}; \cite{lefevre00}; \cite{P00}; \cite{P02}; \cite{conselice03}; \cite{lin04}; \cite{kartaltepe07a}; \cite{lin08}; \cite{lotz08}; \cite{kampczyk}). Even though, $m$ remains poorly 
constrained with $0 \leq m \leq 6$, meaning either no evolution of the merger
rate with cosmic time, or a strong evolution. Part of this large range of values 
can be understood as coming from the different criteria used to identify merger 
candidates, or the photometric 
band used to identify pairs (\cite{bundy}). Furthermore, comparing measurements at low
and high redshifts from different surveys is complicated due to the different selection
functions used. At redshifts $z>0.3$, most pair counts so far have been
performed from a measurement of the number of pairs observed on deep images,
with either a photometric redshift of the galaxies (e.g. \cite{conselice03}; \cite{lotz08}), or a spectroscopic redshift of one of the galaxies in the pair 
(e.g. \cite{P97}; \cite{lefevre00}). 
The effect of contamination by galaxies projected along the line of sight producing false pairs is then estimated from galaxy counts,  
and the observed pair fraction is corrected to get an estimate of the true
pair fraction. As redshift increases, projection effects become increasingly 
important making it difficult to estimate the true pair fraction, creating
a fondamental uncertainty in the measurement of $m$. At $z\sim1$
a galaxy with a luminosity $L_*$ has a 40\% probability to have a galaxy
with a similar magnitude but at a different redshift projected within
an apparent radius of $20 h^{-1} kpc $ (\cite{lefevre00}). 

To overcome these limitations, the most secure method to identify a physical 
pair of galaxies is to obtain a velocity measurement of each galaxy in the pair, 
enabling to identify pairs of galaxies which are most likely to be gravitationnaly bound.
Only recently samples with spectroscopic redshifts for both galaxies in a pair are 
becoming available (\cite{lin07}, \cite{lin08}). 
In this paper we use for the first time a complete redshift survey to $z \sim 1$
and as faint as $I_{AB}=24$
to securely identify pairs with {\it both} galaxies having a
spectroscopic redshift. We use the VIMOS VLT Deep Survey (VVDS) (\cite{lefevre05a}), 
to search for galaxy pairs and to derive the
pair fraction and the merger rate evolution.  We present the galaxy sample
and the methodology to build a pair sample in Section 2, we derive the pair
fraction evolution in Section 3, and we examine the spectrophotometric
properties of galaxies in pairs in Section 4. We compute the merger 
rate in Section 5. We evaluate the fraction of the stellar mass
involved in mergers since $z\sim1$ in Section 6, and conclude in Section 7. 
We adopt a $H_0=70 km s^{-1} Mpc^{-1}$, $\Omega_{\lambda}=0.7$ and $\Omega_{m}=0.3$ 
cosmology throughout this work and magnitudes are given in the AB system.


\section{Identification of galaxy pairs}

   \subsection{VVDS overview}
We use the deep sample from the VIMOS VLT Deep Survey on the 0216-04 field. 
Data have been obtained with the Visible Multi Object Spectrograph (VIMOS) on the 
ESO-VLT UT3 (\cite{lefevre03}). A total of 9842 objects have been observed in the VVDS-Deep field over a total area of $\sim 0.5\ deg^2$, 
selected solely on the basis of apparent magnitude $17.5 \leq I_{AB} \leq 24$. The mean redshift of the sample is $z=0.76$.
The velocity measurement of each galaxy redshift has an accuracy of $\sim 276 km/s$ (Le F\`evre, 2005a).
 A strategy of multiple spectrograph passes have been used (\cite{bottini}), defining areas where targets have been randomly selected in four separate observations, and another area where two independent observations have been performed, leading to an effective random sampling of the galaxy population
of respectively $\sim 35\%$ and $\sim 20\%$ for each of these two areas (with respectively $S_{4p}=0.17\ deg^2$ and $S_{2p}=0.32\ deg^2$). 
We use a catalogue which contains 6464 objects in an effective area of 
$\sim 0.5 \ deg^2$, using only the most secure redshifts, 
i.e. quality flags 2,3,4  and 9 for primary targets and 22, 23, 24 and 29 for secondary 
targets. Flags 2 ,3 ,4  correspond to redshifts measured with a confidence level of 80\%, 
95\% and 100\%, respectively, and flag 9 indicates spectra with single emission line
(see \cite{lefevre05a} for details).

   \subsection{Selection of pairs in the VVDS}

We have identified pairs in two ways. First,  we have searched in the main VVDS catalogue 
to find pairs of galaxies close in separation perpendicular to the plane of the sky 
using the angular distance at the redshift of the pair, and close in velocity  
along the line of sight as derived from the redshift measurements.
Secondly, we have visually examined the 2D spectra to identify secondary objects close to a primary VVDS target
which have escaped the automated spectra detection algorithm (\cite{scodeggio}) because
their angular proximity to the main target along the slit creates
a blend of the two spectra at the faint isophotes used for detection. 
We looked for evidence for 2 continuum 
traces next to each other, with a clear separation of the objects profile along the
slit. The 1D spectrum of the companion was then extracted and its redshift measured
using the cross-correlation with templates as done for the main VVDS sample , and was assigned a flag 3X, with X following the flag
nomenclature of the survey as described in Section 2.1.
We then search in the parent VVDS imaging catalogue for 
the object responsible for the secondary trace, providing its sky coordinates, the magnitudes
and colors. In case the photometric catalogue did not identify the companion also because
of blending, ugri and z images from the CFHTLS 
(http://www.cfht.hawaii.edu/Science/CFHLS/) 
were examined and the multi-band 
photometry of the companion was performed using flux extraction in image areas isolating 
the object.  This process concerned mainly objects with separations
$1 \leq \theta \leq 2$ arcseconds.

Our final catalogue contains all primary target galaxies with 
secure redshift measurements (flags 2, 3, 4, 9) and $I_{AB} \leq 24$ (6287 objects), all secondary
target galaxies identified by the automated spectra extraction program (flags 22, 23, 24, 29) (160 objects) 
and all the companions identified through our visual examination of 2D spectra
(flags 32, 33, 34, 39) (17 objects).

To create the pair catalogue, we first compute two quantities: 
the projected separation $r_p$ and the line-of-sight velocity difference $\Delta v$. 
For a pair of galaxies with redshift $z_i$ and $z_j$ and an angular separation 
$\theta$ these parameters are given by :
\begin{eqnarray}
r_p &=& \theta \times\ d_{A}(z_{m}) ,\ \mathrm{where} \ \ z_m = \frac{z_i + z_j}{2}, \\
\Delta v &=& c\ \frac{| z_i - z_j |}{1+z_m}, \  \nonumber
\end{eqnarray}
where $d_A(z_m)$ is the angular diameter distance at the mean pair redshift $z_m$, 
and $c$ the speed of light.

\begin{figure}[htbp!]
\resizebox{!}{5.5cm}{\scalebox{1}{\input{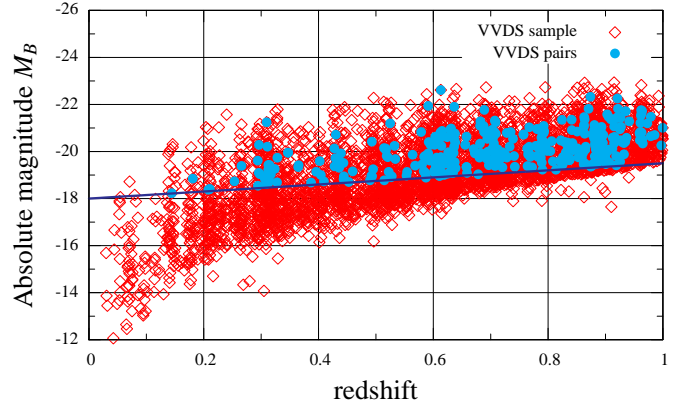}}}
 \caption{Absolute B band magnitude distribution versus redshift of the primary galaxies in a pair with $r_p \leq 100h^{-1} kpc$ and $\Delta v \leq 500 km/s$ (filled symbols) compared to the underlying population of galaxies (open symbols). The line indicates the limit in absolute magnitude used to identify pairs with $M_B \leq -18-Q(z)$.}
 \label{Mag}
 \end{figure}

The $I_{AB} \leq 24$ selection of the VVDS implies that galaxies in the sample 
have an absolute magnitude $M_B \leq -19.11$ at $z = 1$.
We are missing pairs for which one member of the pair is fainter than this limit, for which  we
need to apply a completeness correction as described in Section 2.4. 
From the luminosity function of the complete VVDS sample, we know that the characteristic luminosity $M^{*}$ in B-band evolves with redshift as $Q(z)= 1.11 \times z$ (\cite{ilbert05}). We have therefore applied a magnitude evolution $M_B = -18 - Q(z)$ to our absolute magnitude cutoff when looking for pairs (see Figure~\ref{Mag}).

 \subsection{The VVDS pair catalogue}

\begin{figure*}[htbp]
 \begin{center}
     \resizebox{!}{8cm}{\scalebox{1}{\input{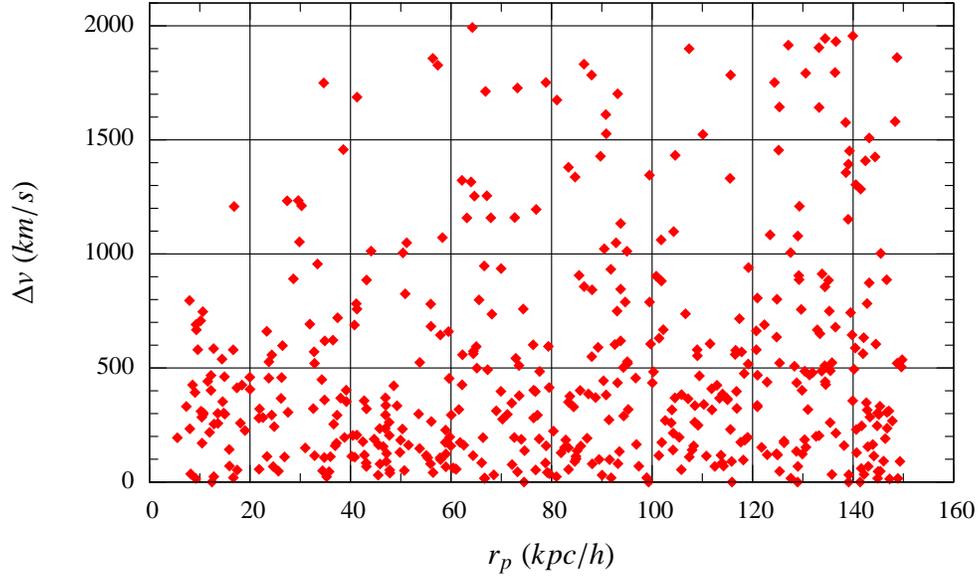}}}\\
    \caption{The line-of-sight velocity difference $\Delta v$ as a function the projected separation $r_p$ for all pairs with $r_p \leq 150h^{-1} kpc$ and\ \  $\Delta M_B^{max} = 1.5$ mag.}
    \label{dv_vs_dr}
 \end{center}
    \end{figure*}

\begin{figure*}[htbp]
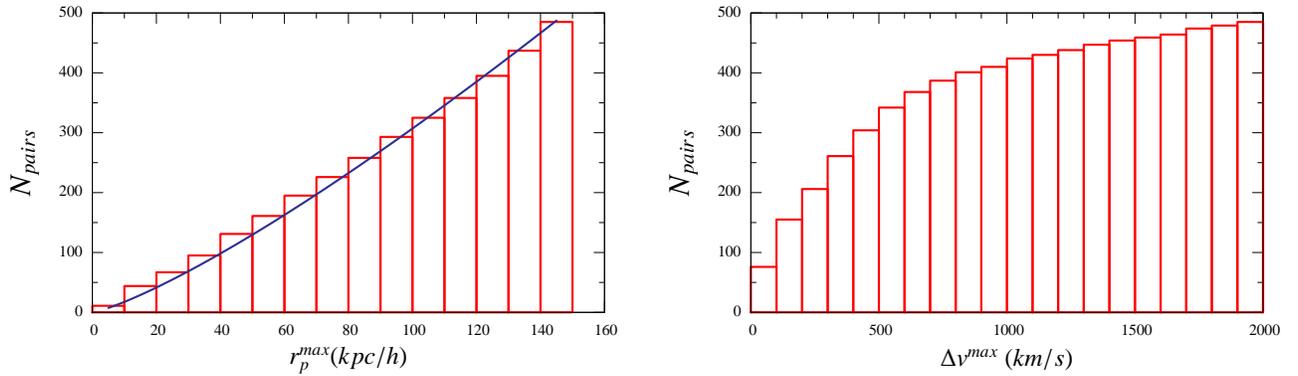

\begin{center}
 \begin{tabular}{c c}
  \resizebox{!}{5cm}{\scalebox{1}{\input{fig3-1.tex}}} & \resizebox{!}{5cm}{\scalebox{1}{\input{fig3-2.tex}}}\\
 
\end{tabular}
\caption{Left : Number of pairs as a function of $r_p^{max}$ for all selected pairs with $\Delta v \leq\ 2000\ km/s\ \ $\ and\ \  $\Delta M_B \leq\ 1.5$ with $M_B \leq \ -18 - Q(z)$. The line represents the best fit to the number of pairs with $N_{pairs}$ $\propto r_p ^{1.24}$, comparable to the expectation $\propto r_p ^{1.3}$ from the angular two-point correlation function. Right : Number of pairs as a function of $\Delta v^{max} $ for all selected pairs with $r_p \leq 150h^{-1}\ kpc$\ and\ \  $\Delta M_B \leq\ 1.5$ with $M_B \leq \ -18 - Q(z)$. }

\label{car-pairs}
\end{center}
\end{figure*}

Using equation (1), we have identified 702 simple pairs and 190 triplets with $r_p^{max}= 150h^{-1} kpc$ and
\ \ $\Delta v^{max} = 2000\ km/s$. To select major mergers, we have imposed an absolute magnitude difference 
between the two members of a pair in the B band of $\Delta M_B \leq 1.5$ mag (see Figure~\ref{dv_vs_dr}).
On the left panel of Figure~\ref{car-pairs} we present the number of pairs with
$\Delta v^{max} = 2000\ km/s\ \ $ and  $\Delta M_B \leq\ 1.5$\ as a function of $r_p$ and on the right panel, the number of pairs with $r_p^{max} = 150h^{-1} kpc\ \ $ and $\Delta M_B \leq\ 1.5$\ as a function of $\Delta v$. 

The two-point correlation function $\xi(r)$ describes the excess probability 
of finding a galaxy at distance r  from a galaxy selected at random over that expected in a uniform, random distribution. This function is usually parametrised by a power law with correlation length $r_0$ : $\xi(r)= (r/r_0)^{-\gamma}$. Integration over this function yields a number of pairs that varies as $r_p^{3-\gamma}$ (\cite{P02}). Using the mean slope $\gamma \sim 1.7$ of the correlation function found in the VVDS  (\cite{lefevre05b}), we expect an increase of the number of pairs $\sim r_p^{1.3}$. This is in good agreement with our pair counts which gives a slope of $\sim 1.24$ as shown in the left panel of Figure~\ref{car-pairs}.

We identify the number of pairs as a function of separations $r_p^{max}$ and  $\Delta v^{max}$
in Table~\ref{table_Npaires}. For $r_p \leq 20h^{-1}\ kpc$, $\Delta v \leq 500\ km/s$ and imposing at least one of the pair members to have $M_B \leq \ -18 - Q(z)$, we have a total of 36 pairs. The fraction of close pairs added by the visual examination of the 
2D spectra is $\sim 10\%$ for pairs with a projected separation less than 2 arcseconds.

\begin{table}[ht]
\center
\caption{Number of pairs with $\Delta M_B^{max} = 1.5$ and $M_B < -18 - Q(z)$. In brackets the number of triplets is given.}
\label{table_Npaires}
\centering
\begin{tabular}{c c c c c}

  & $20h^{-1}\ kpc$  & $30h^{-1}\ kpc$ & $50h^{-1}\ kpc$ & $100h^{-1}\ kpc$\\ \hline\hline \
\\
 500 km/s & 36 (0) & 51 (0) & 102 (6) & 202 (22)\\  
1000 km/s & 48 (0) & 68 (0) & 131 (10) & 267 (32)\\ 
2000 km/s & 50 (0) & 73 (0) & 143 (13) & 314 (46)\\ 
\end{tabular} 
\end{table}

\begin{figure*}[htbf]
\centering
\caption{Postage stamps (6'' $\times$ 6'') and spectra of our 36 VVDS pairs selected in the bright $M_B (z=0) \leq -18$ sample for galaxies with $r_p^{max}=20h^{-1}\ kpc$, $\Delta v^{max}=500\ km/s$ and $\Delta M_B \leq 1.5$. Spectra can be found either at $http://cencosw.oamp.fr/VVDS/VVDS\_DEEP.html$ or $http://vizier.u-strasbg.fr/$. }
  \resizebox{!}{25cm}{\includegraphics{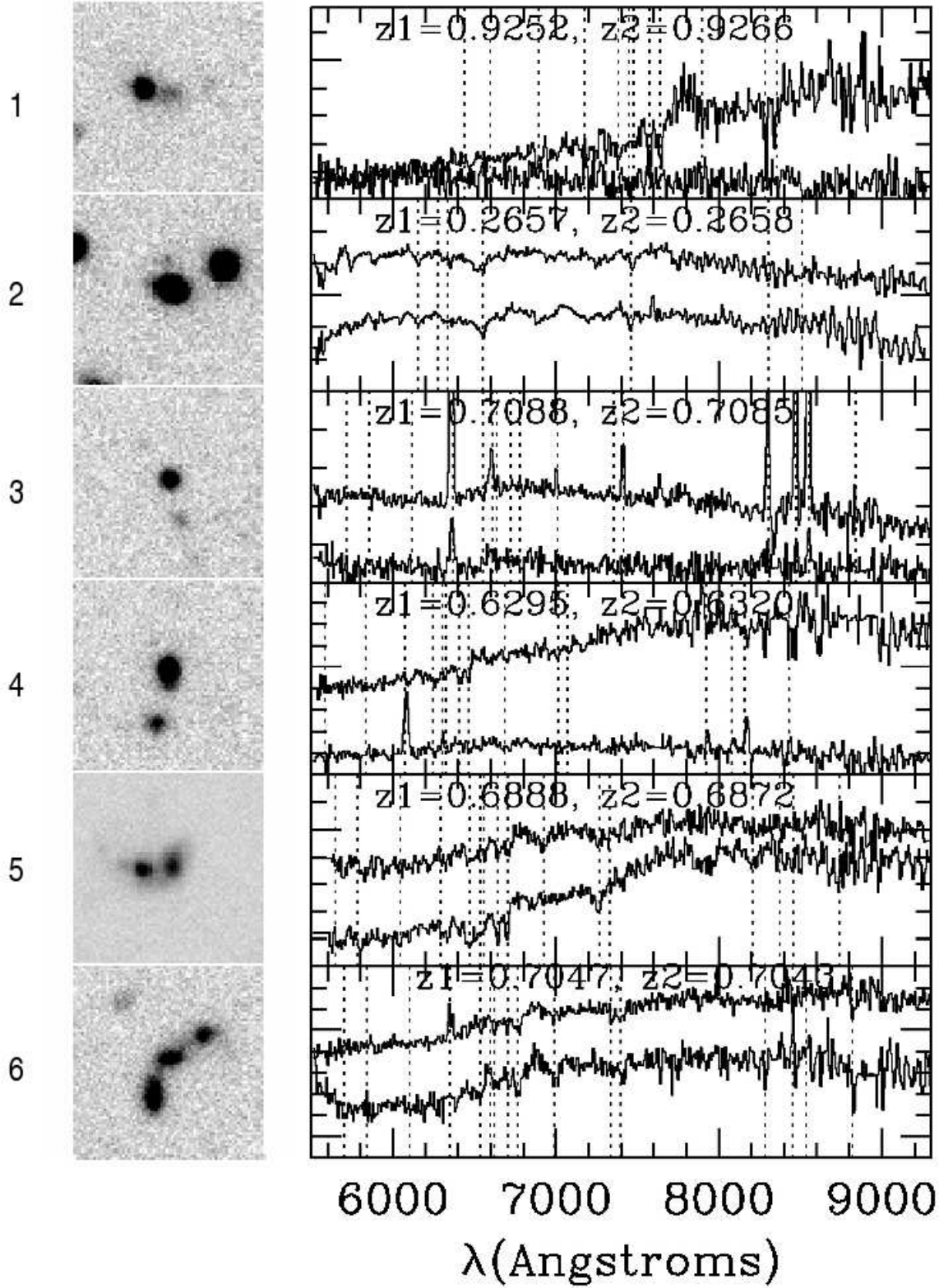}}
\label{stamps}
\end{figure*}

We give the list of all 36 pairs with $r_p \leq 20h^{-1}\ kpc$, $\Delta v \leq 500\ km/s$, $\Delta M_B \leq 1.5$ and $M_B \leq \ -18 - Q(z)$ 
in Table~\ref{list} and we display the postage stamps and spectra of each of these pairs in Figure~\ref{stamps}.

\begin{table*}[ht]
\center
\caption{List of the 36 spectroscopic pairs with $r_p^{max}= 20h^{-1} kpc$, $\ \Delta v^{max} = 500\ km/s\ \ $\ and\ \  $\Delta M_B^{max} = 1.5$ mag selected in the bright $M_B(z=0) \leq -18$ sample. Id's beginning by "p" are manually extracted. Pair numbers can be use to retrieve postage stamps and spectra in Figure~\ref{stamps}. R.A.(2000) and Dec.(2000) are given for the first galaxy.}
\label{list}
\centering
\begin{tabular}{c|c c c c c c c c c c c}
\hline \hline \\
Pair number & Id1 & Id2 & R.A.(2000) & Dec.(2000) & $z1$ & $z2$ & $z_{mean}$ & $r_p\ ( h^{-1} kpc )$ & $\Delta v\ (km/s)$ & $\Delta M_B$ & $\theta\ (")$\\ \hline\hline \\
1 & 020236244 & 020236318 & 36.533573 & -4.483242 & 0.9252 & 0.9266 & 0.9259 &  12.0 & 217.9 &  0.16 &   2.2\\ 
2 & 020260617 & 020260636 & 36.720883 & -4.429412 & 0.2657 & 0.2658 & 0.2657 &  12.8 &  23.7 &  0.02 &   4.5\\ \
3 & 020260559 & 020260085 & 36.655374 & -4.427911 & 0.7088 & 0.7085 & 0.7087 &  17.4 &  52.6 &  1.44 &   3.5\\ \
4 & 020274782 & 020273998 & 36.521896 & -4.396756 & 0.6295 & 0.6320 & 0.6308 &  20.0 & 459.6 &  1.04 &   4.2\\ \
5 & 020314240 & 020314107 & 36.585748 & -4.309167 & 0.6888 & 0.6872 & 0.6880 &  10.4 & 284.2 &  0.92 &   2.1\\ \
6 & 020461143 & 020461037 & 36.699123 & -4.399960 & 0.7047 & 0.7043 & 0.7045 &  15.9 &  70.4 &  0.22 &   3.2\\ 
7 & 020199214 & 020199508 & 36.277674 & -4.557759 & 0.9126 & 0.9107 & 0.9116 &  15.0 & 298.0 &  0.59 &   2.7\\ \
8 & 020205594 & 020204675 & 36.714690 & -4.545729 & 0.6309 & 0.6323 & 0.6316 &  13.6 & 257.2 &  0.11 &   2.8\\ \
9 & 020208200 & 020207985 & 36.647773 & -4.538258 & 0.6946 & 0.6929 & 0.6937 &  13.6 & 300.9 &  0.98 &   2.7\\ \
10 & 020231154 & 020230801 & 36.518347 & -4.493709 & 0.9265 & 0.9250 & 0.9257 &   8.0 & 233.5 &  0.73 &   1.5\\ \
11 & 020234145 & 020234032 & 36.685092 & -4.486811 & 0.8859 & 0.8885 & 0.8872 &  17.4 & 413.0 &  0.95 &   3.2\\ \
12 & 020236142 & 020235785 & 36.537859 & -4.482770 & 0.6206 & 0.6229 & 0.6218 &  18.4 & 425.2 &  1.24 &   3.9\\ \
13 & 020141586 & 020141929 & 36.288443 & -4.692461 & 0.8310 & 0.8336 & 0.8323 &   8.6 & 425.4 &  0.70 &   1.6\\ \
14 & 020161356 & p020161356 & 36.461850 & -4.647027 & 0.9349 & 0.9360 & 0.9355 &  10.5 & 170.4 &  0.43 &   1.9\\ \
15 & 020162148 & 020162920 & 36.528063 & -4.644043 & 0.6815 & 0.6817 & 0.6816 &   8.2 &  35.7 &  0.03 &   1.7\\ \
16 & 020170414 & 020170218 & 36.436325 & -4.628108 & 0.3646 & 0.3625 & 0.3635 &  14.9 & 461.7 &  0.46 &   4.2\\ \
17 & 020182684 & 020182811 & 36.465562 & -4.597417 & 0.7022 & 0.7002 & 0.7012 &  14.6 & 352.4 &  0.24 &   2.9\\ \
18 & 020198752 & 020198370 & 36.958101 & -4.559067 & 0.9366 & 0.9366 & 0.9366 &  12.4 &   0.0 &  1.22 &   2.3\\ \
19 & 020281203 & 020281920 & 36.931009 & -4.381017 & 0.9004 & 0.9032 & 0.9018 &  11.6 & 441.4 &  0.99 &   2.1\\ \
20 & 020383500 & 020384409 & 36.753843 & -4.157210 & 0.5639 & 0.5640 & 0.5639 &  16.7 &  19.2 &  0.64 &   3.7\\ \
21 & 020323722 & 020323591 & 36.583688 & -4.286744 & 0.9270 & 0.9240 & 0.9255 &  12.2 & 467.1 &  1.21 &   2.2\\ \
22 & 020336174 & p020336174 & 36.883994 & -4.259516 & 0.6999 & 0.6988 & 0.6994 &   5.5 & 194.1 &  1.12 &   1.1\\ \
23 & 020226763 & 020226762 & 36.771212 & -4.502420 & 0.5683 & 0.5662 & 0.5673 &  12.2 & 401.7 &  0.79 &   2.7\\ \
24 & 020113570 & 020113267 & 36.553019 & -4.756721 & 0.7199 & 0.7218 & 0.7208 &   7.3 & 331.0 &  0.84 &   1.4\\ \
25 & 020462322 & 020462321 & 36.730651 & -4.378359 & 0.9322 & 0.9323 & 0.9323 &   9.2 &  15.5 &  0.53 &   1.7\\ \
26 & 020462055 & 020462033 & 36.636971 & -4.383065 & 0.9186 & 0.9205 & 0.9195 &  10.7 & 296.7 &  0.01 &   1.9\\ \
27 & 020461384 & 020461394 & 36.749187 & -4.396416 & 1.1846 & 1.1824 & 1.1835 &  14.8 & 302.1 &  0.57 &   2.5\\ \
28 & 020164724 & 020164374 & 36.595220 & -4.638303 & 0.6059 & 0.6038 & 0.6048 &   9.1 & 392.3 &  0.46 &   1.9\\ \
29 & 020214961 & 020215062 & 36.611008 & -4.520846 & 0.7420 & 0.7405 & 0.7412 &  18.1 & 258.3 &  0.12 &   3.5\\ \
30 & 020255699 & 020255847 & 36.649314 & -4.438386 & 0.8854 & 0.8870 & 0.8862 &  12.8 & 254.3 &  0.27 &   2.4\\ \
31 & 020294680 & 020295035 & 36.914281 & -4.349939 & 0.7265 & 0.7252 & 0.7258 &  19.0 & 225.8 &  1.09 &   3.7\\ \
32 & 020172440 & 020172473 & 36.815309 & -4.620937 & 0.5437 & 0.5429 & 0.5433 &  18.2 & 155.4 &  0.43 &   4.1\\ \
33 & 020158576 & 020158574 & 36.474773 & -4.652988 & 0.6810 & 0.6818 & 0.6814 &  14.0 & 142.6 &  0.10 &   2.8\\ \
34 & 020196960 & 020196959 & 36.858070 & -4.562418 & 1.2729 & 1.2706 & 1.2717 &  10.7 & 303.5 &  1.26 &   1.8\\ \
35 & 020231368 & 020231271 & 36.474256 & -4.492534 & 1.0984 & 1.0994 & 1.0989 &  15.9 & 142.8 &  1.42 &   2.8\\ \
36 & 020469171 & p020469171 & 36.668148 & -4.260920 & 0.8381 & 0.8362 & 0.8371 &  10.2 & 310.0 &  1.39 &   1.9\\ \hline

\end{tabular} 
\end{table*}

\begin{figure*}[htbf]
\centering
  \resizebox{!}{25cm}{\includegraphics{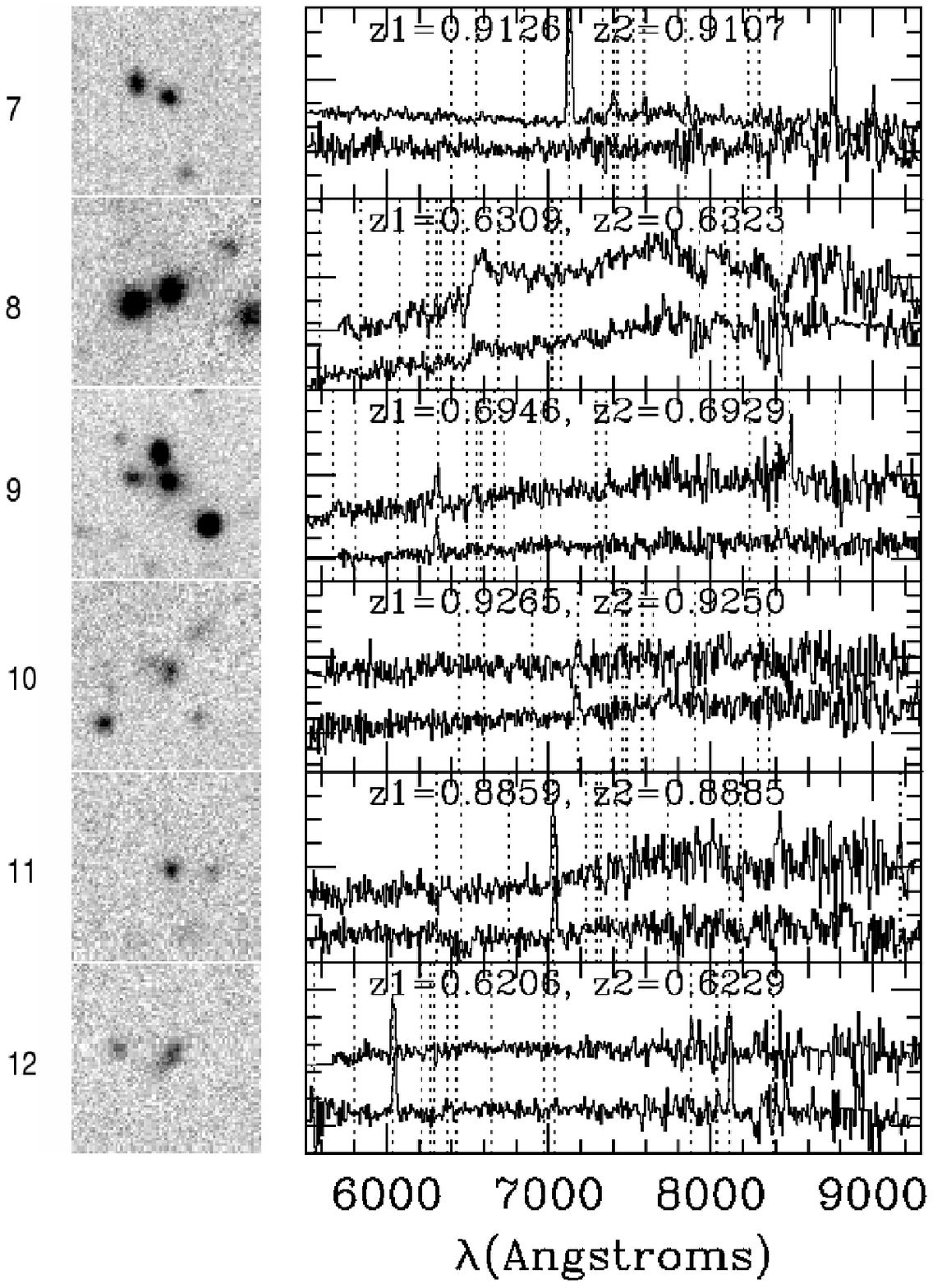}}
\end{figure*}

\clearpage

 \begin{figure*}[htbf]
\centering
  \resizebox{!}{25cm}{\includegraphics{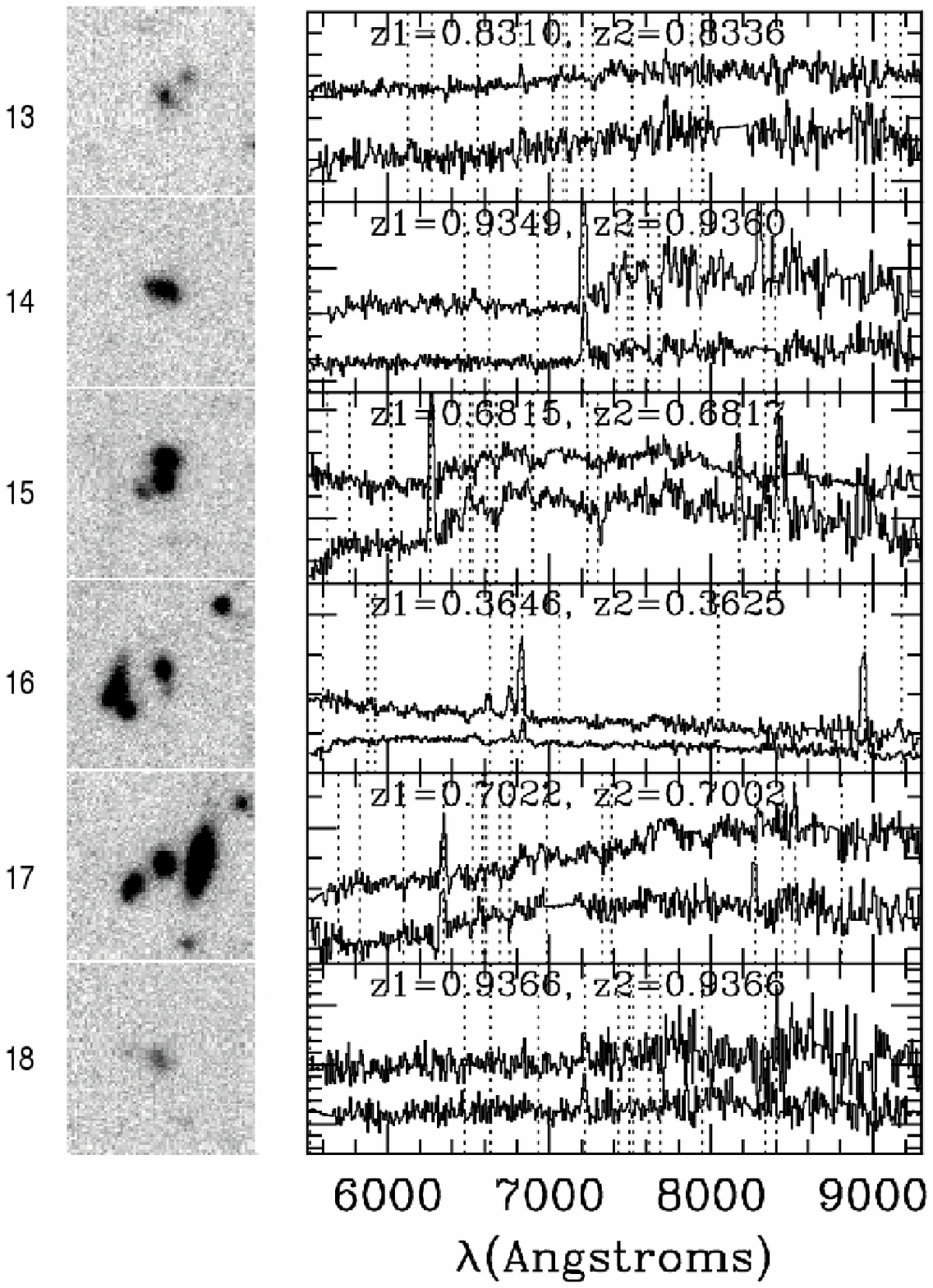}}
\end{figure*}

\clearpage

 \begin{figure*}[htbf]
\centering
  \resizebox{!}{25cm}{\includegraphics{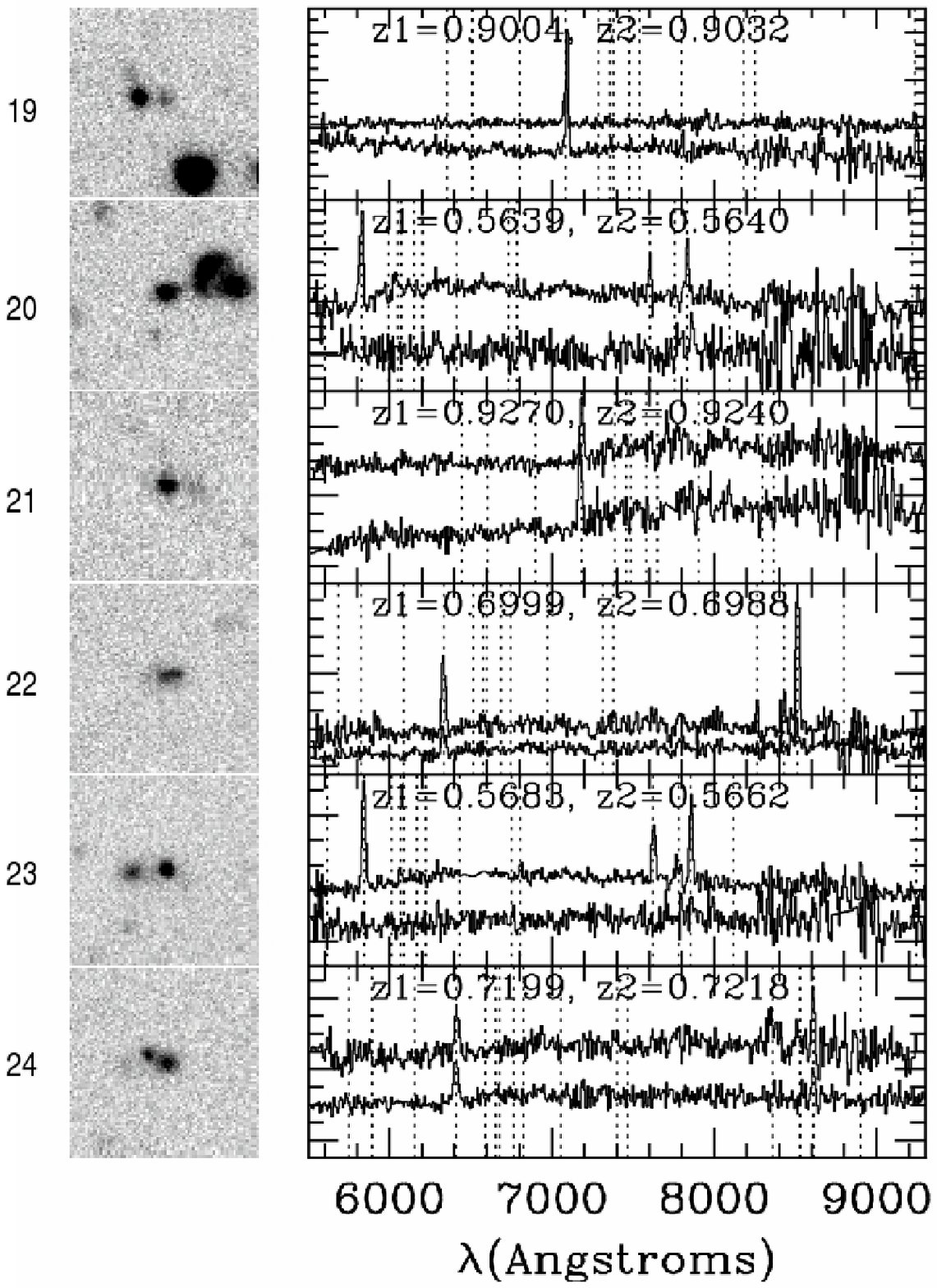}}
\end{figure*}

\clearpage

\begin{figure*}[htbf]
\centering
  \resizebox{!}{25cm}{\includegraphics{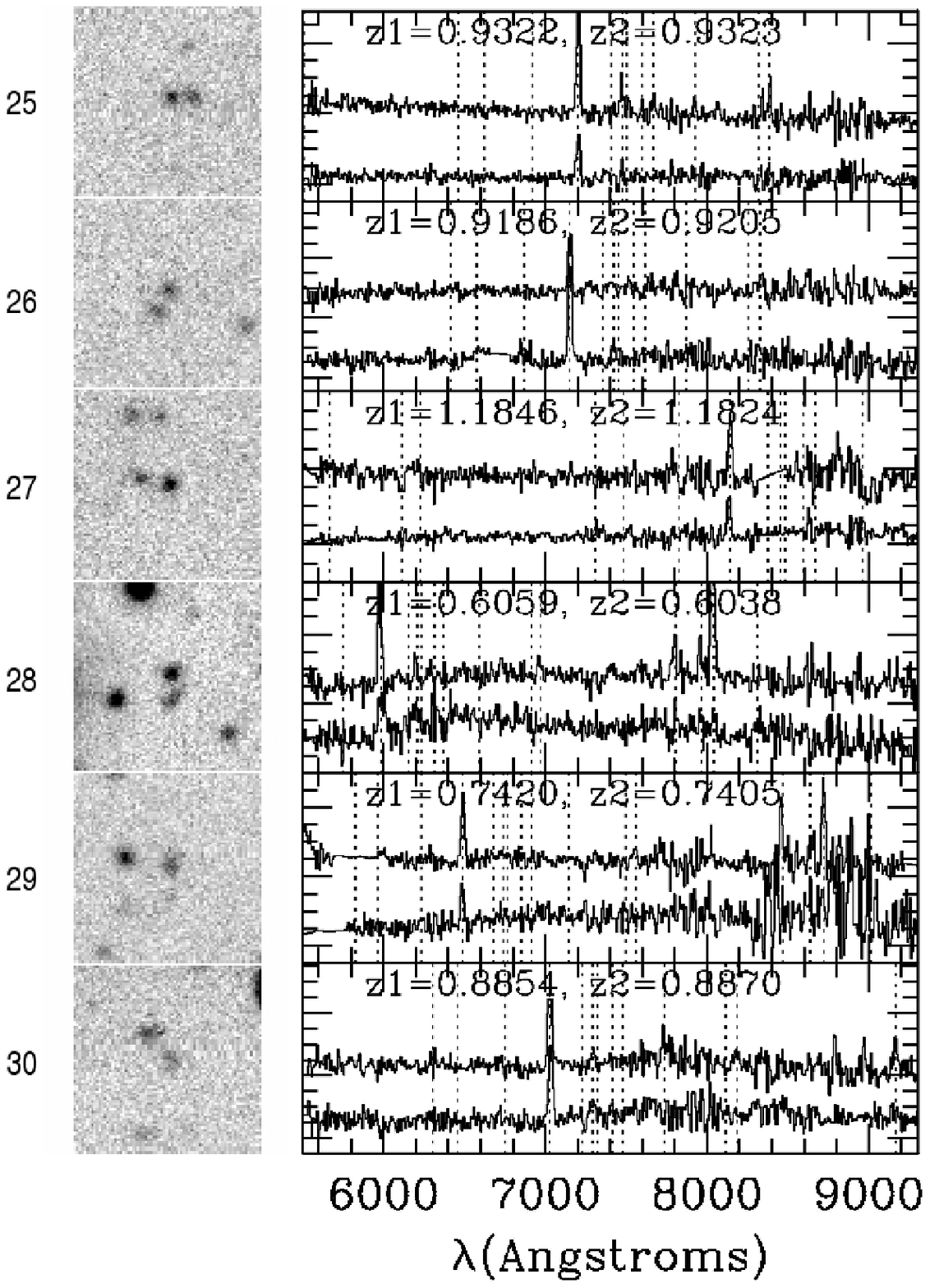}}
\end{figure*}

\clearpage

\begin{figure*}[htbf]
\centering
  \resizebox{!}{25cm}{\includegraphics{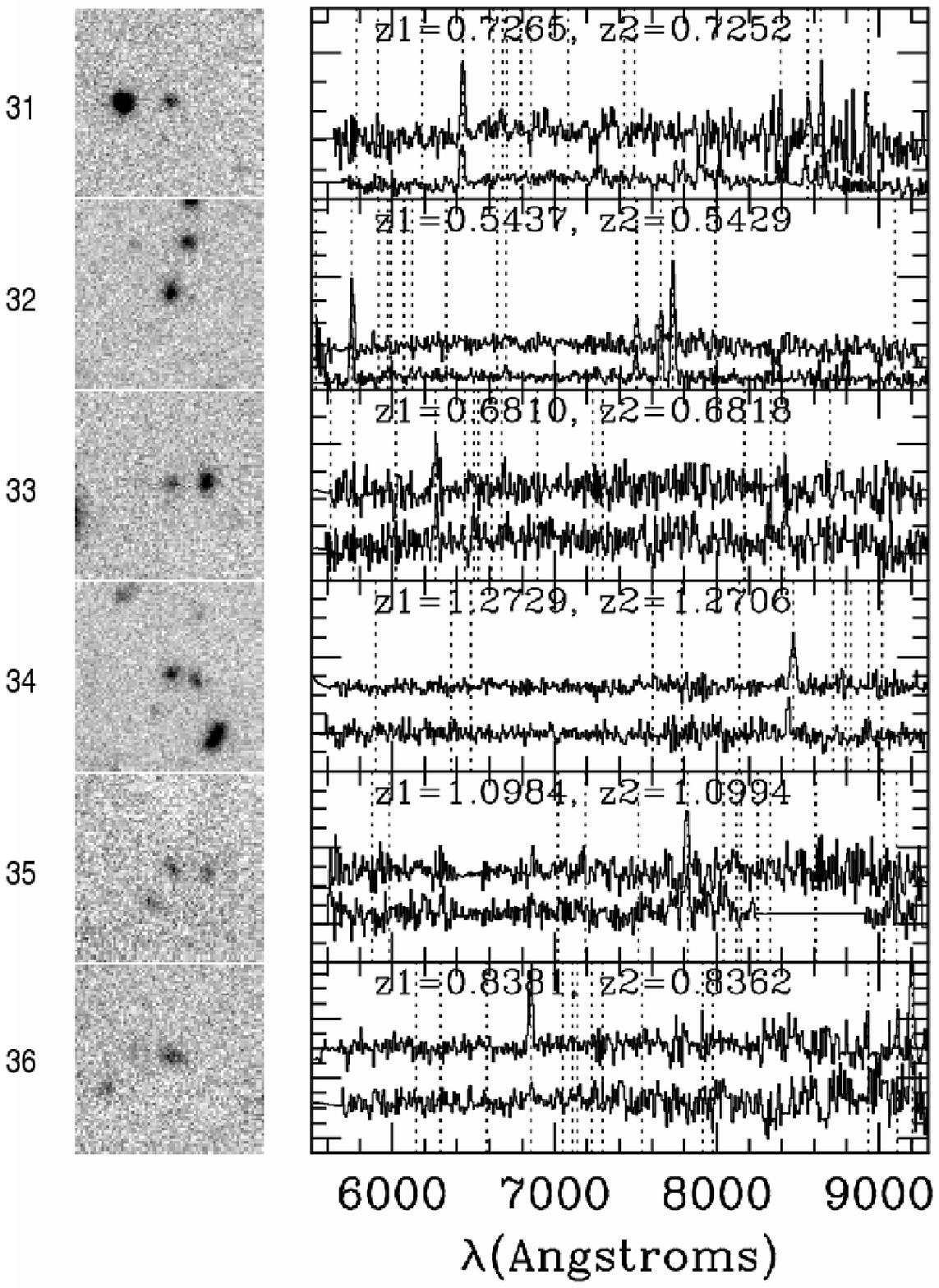}}
\end{figure*}

\clearpage

   \subsection{Accounting for selection effects}

To compute the total number of true pairs, we need to correct for three basic 
effects imposed by the VVDS selection function: 
\begin{enumerate}
 \item{the limiting magnitude
$I_{AB}=24$ which imposes a loss of faint companions when we search for major
mergers with $\Delta M_B \leq 1.5$.}
 \item{the spatial sampling rate and the 
spectroscopic success rate in measuring redshifts.}
 \item{the loss of 
pairs at small separations because of the ground based seeing limitation
of the observations.}
\end{enumerate}

The spectroscopic targets have been selected on the basis of the only magnitude criterion 
$17.5 \leq I_{AB} \leq 24$. Therefore, we miss companions which have an absolute 
magnitude fainter than imposed by the $I_{AB}=24$ cutoff and the $\Delta M_B=1.5$ magnitude
difference, artificially lowering the number of pairs. To take this into account, we
compute for each galaxy a weight $\omega_{mag}(M_B,z)$ using the ratio between the comoving number densities above  and below the magnitude cutoff (\cite{ilbert05}). For each galaxy, we derive $M_{sup}^{i}=M_B^{i}+\Delta M_B$ which corresponds to the maximum absolute magnitude when searching for a companion and $M_{sel}^{i}(z)$ which corresponds to the survey limit $I=24$ in the absolute B band at the given galaxy redshift. We then assign a weight for each galaxy: 
$$\omega_{mag}^{i}(M_B,z) = \left\{\begin{array}{ll}
 1 & \mbox{if $M_{sup}^{i} \leq M_{sel}^{i}$}\\
\\
 \frac{\displaystyle \int_{-\infty}^{M_{sup}^{i}} \Phi(M) dM}{\displaystyle \int_{-\infty}^{M_{sel}^{i}} \Phi(M) dM} & \mbox{if $M_{sup}^{i} > M_{sel}^{i}$.}
 \end{array}\right.$$

We combine these weights in each pair $k$ as $\omega_{p,mag}^{k}=\omega_{mag}^{i} \times \omega_{mag}^{j}$ where $\omega_{mag}^{i}$ and $\omega_{mag}^{j}$ are the weights of each galaxie in the pair.

Since 25\% of the field has been spectroscopicaly observed and the redshifts are not measured with 100\% certainty, 
we must correct for the VVDS sampling rate and redshift success rate.
These have been very well constrained (see \cite{ilbert05}) resulting in the Target Sampling Rate (TSR) and the 
Spectroscopic Success Rate (SSR) computed as a function of redshift.
 The SSR has been assumed independent of the galaxy type, as demonstrated to be true up to $z \sim 1$ in Zucca et al. (2006). We therefore introduce the weight $\omega_{comp}^{i} (z)$. For each galaxy, we have the information on its redshift, its apparent magnitude $I_{AB}$, its spectroscopic flag and its spatial flag 
(whether the galaxy is on the field with four passes or two passes).
We derive the completness weight as follows.
$$\omega_{comp}^{i}(z)=\left(\frac{\displaystyle N_{g,spectro}^{sel}(z)}{\displaystyle N_{g,photo}^{sel}(z)}\right)^{-1} ,$$
where $N_{g,spectro}^{sel}$ is the number of secure spectroscopic flag galaxies in the spectroscopic catalogue, and where  $N_{g,photo}^{sel}$ is the number of galaxies in the photometric catalogue. These two last values are estimated within the same redshift, $I$-band magnitude and N-pass area ranges based on the $z$, $I_{AB}$ and N-pass area values of the galaxy $i$.
For the photometric sample, we use the photometric redshifts of Ilbert et al. (2005). Each pair $k$ is therefore assigned with $\omega_{p,comp}^{k}=\omega_{comp}^{i} \times \omega_{comp}^{j}$ where $\omega_{comp}^{i}$ and $\omega_{comp}^{j}$ are the completeness weights of each galaxie in the pair.
\\

The last correction we need to apply results from the observations which have been performed
under a typical ground based seeing of 1 arcsecond. We correct for the increasing incompleteness
to target both components of close pairs as the separation between them
is getting smaller. Assuming a clustered 
distribution of galaxies, the number of galaxy pairs should be a 
monotonically decreasing function of the pair separation.
However, pairs are under-counted for separations
$\theta \leq 2$ arcseconds because of the seeing effects.

\begin{figure}[h!]
 \begin{center}
     \resizebox{!}{5.5cm}{\scalebox{1}{\input{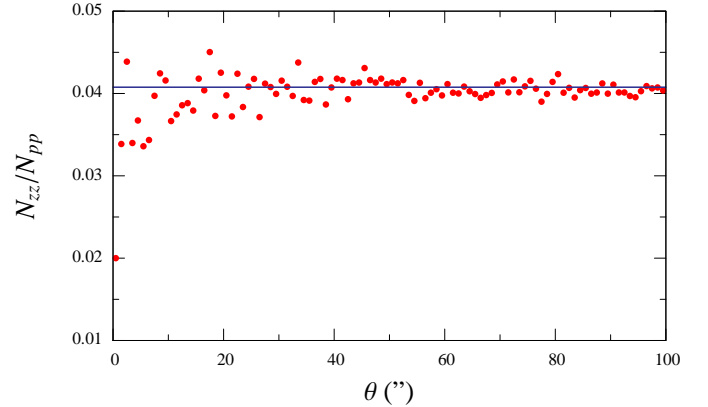}}}\\
    \caption{Spectroscopic completeness as a function the angular pair separations. The line is the fit used to derive 
the mean correcting factor $a$ which corresponds roughly to the square of the completeness.}
    \label{correction}
 \end{center}
    \end{figure}

We derive the ratio $r(\theta)$ between the observed pair count in the spectroscopic catalogue, $N_{zz}$, over the observed pair count in the photometric catalogue, $N_{pp}$, as a function of the angular separation (see Figure~\ref{correction}). We apply a weight $\omega_{\theta}^{k}$ on each pair $k$ using the ratio :
\\

$$\omega_{\theta}^{k}= \frac{\displaystyle a}{r(\displaystyle \theta_k)} ,$$ where the mean ratio $a$ is the probability to randomly select a pair, obtained at large separations. This ratio is close to the squared mean target sampling rate ($\sim 20.2 \% ^2$). For large separations ($\theta > 50 ''$), $r(\theta) \sim a$ but at small separations $r(\theta) < a$ because of the artificial decrease of pairs due to seeing effects.
\\

The corrected number of galaxies $N_{g}^{corr}$ in each redshift bin is then:

\begin{eqnarray}
 N_{g}^{corr}(z) = \sum_{i=1}^{N_{g,obs}} \omega_{comp}^{i} \times \omega_{mag}^{i}.
\end{eqnarray}

The total number of pairs $N_p^{corr}$ is therefore computed as :
\begin{eqnarray}
N_p^{corr}(z) =  \sum_{k=1}^{N_{p,obs}} \omega_{p,comp}^{k} \times \omega_{p,mag}^{k} \times \omega_{\theta}^{k} ,
\end{eqnarray}
where $N_{g,obs}$ and $N_{p,obs}$ are the observed number of galaxies and pairs in the spectroscopic catalogue.

\section{Evolution of the pair fraction with redshift}

   \subsection{Pair fraction evolution using VVDS data}
   \label{3.1}

We give the total number of identified pairs as a function of the two 
separations criteria in Table 1 for the adopted $\Delta M_B \leq 1.5$ magnitude difference. 
We use equations (2) \& (3) to compute the pair fraction $f_p(z)$ in each redshift bin as follows: 
\begin{eqnarray}
f_p(z)=\frac{N_{p}^{corr}(z)}{N_{g}^{corr}(z)}.
\end{eqnarray}
Table~\ref{table_PF} gives values of $f_p(z)$ for different sets of $r_p^{max}$ and $\Delta v^{max}$ derived using the $M_B(z) = -18 -Q(z)$ relation derived for the VVDS sample. 
Using the parameterization $f_p(z) = f_{p}(0) \times (1+z)^m$, we fit the pair fraction measurements to compute the evolution index $m$ and associated
poissonian errors as a function of the line-of-sight and projected transverse separations.
 These values are reported in Table~\ref{m_No}.

\begin{figure}[h!]
 \begin{center}
     \resizebox{!}{5.5cm}{\scalebox{1}{\input{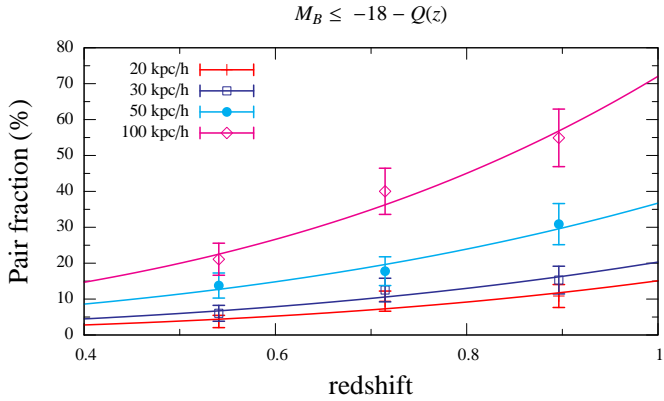}}}\\
    \caption{Evolution of the pair fraction as a function of redshift for different sets of 
$r_p^{max}$ ,\ $\Delta v^{max} =\ 500\ km/s$ and galaxies brighter than $M_B(z) = -18 -Q(z)$.}
    \label{fit-pair_fraction}
 \end{center}
    \end{figure} 

Figure~\ref{fit-pair_fraction} shows 
the evolution of the pair fraction of galaxies with $r_p^{max} = 20,\ 30,\ 50,\ 100 h^{-1}\ kpc$,\ $\Delta_v^{max} = 500\ km/s$ and $\Delta M_{B} \leq 1.5 $. 
A total of $\sim 10.86 \pm 3.20\%$ of galaxies with $M_B \leq -18 - Q(z)$ 
and $r_p^{max} = 20h^{-1} kpc$ are in close pairs at $z \sim 0.9$ compared to $ \sim 3.76 \pm 1.71\%$ at 
$ z \sim 0.5$. This leads to $f_p = (0.57 \pm 0.65 \%) \times (1+z)^{4.73 \pm 2.01}$. 
The fraction of galaxies brighter than $M_B = -18 - Q(z)$ in pairs increases significantly with redshift.

We have investigated the dependency of the pair fraction on the pair separation. 
Increasing the separation of the two members of a pair both in $r_p$ and $\Delta v$, the index $m$
varies from $4.73 \pm 2.01$ to $2.45 \pm 0.11$ when  separations increase from $( 20 h^{-1}\ kpc,\ 500 km/s )$ to $ (100 h^{-1}\ kpc,\ 2000 km/s) $. 

Interestingly, we find a strong dependency on the limiting absolute magnitude of the galaxies in the pairs. Table \ref{table_PF} gives the pair fractions for different redshift, $r_p^{max}$ and $\Delta v^{max}$ using the $M_B(z=0) \leq -18.77$ VVDS sample and Table \ref{m_No} gives the best fit values of $m$ and $f_p (z=0)$. For $r_p^{max} = 20h^{-1}\ kpc$ and $\Delta v^{max} = 500\ km/s$, $m$ decreases from $m=4.73 \pm 2.01$ in the complete faint sample to $m=3.07 \pm 1.68$ for the bright sample with $M_B(z=0) \leq -18.77$, implying a weaker evolution. This trend is seen for any separations from ($20h^{-1}\ kpc$, $500\ km/s$) to ($100h^{-1}\ kpc$, $2000\ km/s$). Here, we show that the pair fraction evolves faster for fainter samples. We will come back to this property in Section \ref{discuss}.

\setlength{\extrarowheight}{4 pt}
\begin{table*}[htbp]
\caption{Pair fraction (in \%) for different sets of separations and redshift, with $M_B \leq -18 - Q(z)$ (faint sample) and with $M_B \leq -18.77 - Q(z)$ (bright sample) using VVDS data. (See Section 3.1)}
\label{table_PF}
\begin{tabular}{|p{1.06cm} c c c c | p{1.06cm} c c c c |}
\hline 
\multicolumn{5}{|c|}{$\mathbf{M_B^{max}(z=0) = -18}$} & \multicolumn{5}{c|}{$\mathbf{M_B^{max}(z=0) = -18.77}$}\\  \hline
\multicolumn{10}{|c|}{$\mathbf{\Delta v \leq 500 km/s}$} \\  \hline
 & $20 h^{-1} kpc$ & $30 h^{-1} kpc$ & $50 h^{-1} kpc$ & $100 h^{-1} kpc$ & & $20 h^{-1} kpc$ & $30 h^{-1} kpc$ & $50 h^{-1} kpc$ & $100 h^{-1} kpc$\\  \hline 
$z=0.54$ & $3.76 \pm 1.71$ & $6.05 \pm 2.21$ & $13.78 \pm 3.50$ & $21.10 \pm 4.46$ & $z=0.51$ & $2.26 \pm 0.52$ & $7.01 \pm 3.00$ & $14.66 \pm 4.54$ & $4.10 \pm 0.40$ \\  
$z=0.71$ & $9.43 \pm 2.80$ & $12.52 \pm 3.28$ & $17.78 \pm 4.00$ & $40.02 \pm 6.43$ & $z=0.70$ & $3.21 \pm 0.77$ & $12.81 \pm 3.65$ & $18.87 \pm 4.57$ & $22.03 \pm 5.75$ \\  
$z=0.90$ & $10.86 \pm 3.20$ & $15.30 \pm 3.86$ & $30.88 \pm 5.73$ & $54.91 \pm 8.01$ & $z=0.90$ & $4.10 \pm 0.40$ & $14.01 \pm 3.70$ & $25.45 \pm 5.20$ & $40.84 \pm 7.25$ \\  
\hline 
\multicolumn{10}{|c|}{$\mathbf{\Delta v \leq 1000 km/s}$} \\  \hline
 & $20 h^{-1} kpc$ & $30 h^{-1} kpc$ & $50 h^{-1} kpc$ & $100 h^{-1} kpc$ & & $20 h^{-1} kpc$ & $30 h^{-1} kpc$ & $50 h^{-1} kpc$ & $100 h^{-1} kpc$\\  \hline 
$z=0.54$ & $6.42 \pm 2.28$ & $8.70 \pm 2.70$ & $17.07 \pm 3.95$ & $32.01 \pm 5.70$ & $z=0.51$ & $7.60 \pm 3.14$ & $9.90 \pm 3.64$ & $17.55 \pm 5.04$ & $33.61 \pm 7.38$\\  
$z=0.71$ & $12.44 \pm 3.27$ & $16.27 \pm 3.81$ & $23.68 \pm 4.72$ & $51.46 \pm 7.48$ & $z=0.70$ & $10.16 \pm 3.20$ & $14.59 \pm 3.93$ & $23.14 \pm 5.15$ & $49.09 \pm 8.12$\\  
$z=0.90$ & $13.45 \pm 3.59$ & $19.69 \pm 4.44$ & $38.73 \pm 6.53$ & $66.65 \pm 9.00$ & $z=0.90$ & $11.29 \pm 3.28$ & $17.37 \pm 4.18$ & $31.29 \pm 5.86$ & $55.22 \pm 8.22$\\  
 \hline 
\multicolumn{10}{|c|}{$\mathbf{\Delta v \leq 2000 km/s}$}\\  \hline
 & $20 h^{-1} kpc$ & $30 h^{-1} kpc$ & $50 h^{-1} kpc$ & $100 h^{-1} kpc$ & & $20 h^{-1} kpc$ & $30 h^{-1} kpc$ & $50 h^{-1} kpc$ & $100 h^{-1} kpc$\\  \hline 
$z=0.54$ & $7.22 \pm 2.44$ & $9.50 \pm 2.84$ & $21.34 \pm 4.49$ & $45.63 \pm 7.06$ & $z=0.51$ & $7.60 \pm 3.14$ & $9.90 \pm 3.64$ & $20.65 \pm 5.53$ & $41.99 \pm 8.44$\\  
$z=0.71$ & $12.44 \pm 3.27$ & $17.89 \pm 4.02$ & $26.09 \pm 4.99$ & $58.05 \pm 8.05$ & $z=0.70$ & $10.16 \pm 3.20$ & $16.45 \pm 4.22$ & $25.91 \pm 5.50$ & $55.17 \pm 8.73$\\  
$z=0.90$ & $14.38 \pm 3.73$ & $20.62 \pm 4.56$ & $40.40 \pm 6.69$ & $75.65 \pm 9.71$ & $z=0.90$ & $12.19 \pm 3.43$ & $18.28 \pm 4.30$ & $32.92 \pm 6.04$ & $62.42 \pm 8.86$\\  
\hline
\end{tabular}
\end{table*}

\begin{table*}[htbp]
\caption{Best fits parameters for $m$ and $f_p (z=0)$ of major mergers as a function of the dynamical parameters in the faint $M_B(z=0) \leq -18$ sample and in the bright $M_B(z=0) \leq -18.77$ one.}
\label{m_No}
\centering
\begin{tabular}{| c c c c c | c c c c c |}
\hline 
\multicolumn{5}{|c|}{$\mathbf{M_B^{max}(z=0) = -18}$} & \multicolumn{5}{c|}{$\mathbf{M_B^{max}(z=0) = -18.77}$}\\ \hline
\multicolumn{10}{|c|}{$\mathbf{\Delta v \leq 500 km/s}$}\\ \hline
 $r_p^{max}$ & $20 h^{-1} kpc$ & $30 h^{-1} kpc$ & $50 h^{-1} kpc$ & $100 h^{-1} kpc$ & $r_p^{max}$ & $20 h^{-1} kpc$ & $30 h^{-1} kpc$ & $50 h^{-1} kpc$ & $100 h^{-1} kpc$\\ \hline
$m$ & $4.73 \pm 2.01$ & $4.24 \pm 1.36$ & $4.07 \pm 0.95$ & $4.46 \pm 0.81$ & $m$ & $3.07 \pm 1.68$ & $3.00 \pm 1.38$ & $2.69 \pm 0.16$ & $3.18 \pm 1.34$\\ 
$f_p (z=0)$ & $0.57 \pm 0.65$ & $1.08 \pm 0.83$ & $2.19 \pm 1.18$ & $3.27 \pm 1.51$ & $f_p (z=0)$ & $1.44 \pm 1.39$ & $2.15 \pm 1.70$ & $4.49 \pm 0.42$ & $6.22 \pm 4.80$\\ 
\hline 
\multicolumn{10}{|c|}{$\mathbf{\Delta v \leq 1000 km/s}$}\\ \hline
 $r_p^{max}$ & $20 h^{-1} kpc$ & $30 h^{-1} kpc$ & $50 h^{-1} kpc$ & $100 h^{-1} kpc$ & $r_p^{max}$ & $20 h^{-1} kpc$ & $30 h^{-1} kpc$ & $50 h^{-1} kpc$ & $100 h^{-1} kpc$\\ \hline 
$m$  & $3.37 \pm 1.52$ & $3.77 \pm 1.13$ & $4.05 \pm 0.55$ & $3.46 \pm 0.54$ & $m$  & $1.81 \pm 0.51$ & $2.57 \pm 0.57$ & $2.80 \pm 0.10$ & $2.25 \pm 0.70$\\ 
$f_p (z=0)$ & $1.66 \pm 1.43$ & $1.86 \pm 1.18$ & $2.84 \pm 0.88$ & $7.46 \pm 2.26$ & $f_p (z=0)$ & $3.62 \pm 1.04$ & $3.41 \pm 1.11$ & $5.15 \pm 0.31$ & $13.4 \pm 5.36$\\ 
\hline 
\multicolumn{10}{|c|}{$\mathbf{\Delta v \leq 2000 km/s}$}\\ \hline
 $r_p^{max}$ & $20 h^{-1} kpc$ & $30 h^{-1} kpc$ & $50 h^{-1} kpc$ & $100 h^{-1} kpc$ & $r_p^{max}$ & $20 h^{-1} kpc$ & $30 h^{-1} kpc$ & $50 h^{-1} kpc$ & $100 h^{-1} kpc$\\ \hline 
$m$ & $3.19 \pm 1.04$ & $3.56 \pm 1.27$ & $3.18 \pm 0.75$ & $2.45 \pm 0.11$ & $m$ & $2.21 \pm 0.29$ & $2.71 \pm 1.08$ & $2.25 \pm 0.06$ & $1.83 \pm 0.40$\\ 
$f_p (z=0)$ & $1.96 \pm 1.15$ & $2.24 \pm 1.61$ & $5.10 \pm 2.14$ & $15.7 \pm 0.98$ & $f_p (z=0)$ & $2.98 \pm 0.49$ & $3.36 \pm 2.08$ & $7.72 \pm 0.27$ & $19.6 \pm 4.43$\\ 
\hline 
\end{tabular} 
\end{table*}

\subsection{Constraints combining low redshift pair fraction with VVDS estimates}
\label{3.2}

To better constrain the evolutionary parameters, the comparison of high redshift data to the local value of the pair fraction is important. Patton et al. (2000) derived the pair fraction in a sample of 5426 galaxies in the SSRS2 redshift survey. Using close ($5\ \leq r_p \leq 20h^{-1}\ kpc$) dynamical ($\Delta v \leq 500 km/s$) pairs, they found $f_p (-21 \leq M_B - 5\ log\ h \leq -18) = 2.26 \pm 0.52\%$ at $z = 0.015$. We also compare our data to results from the CNOC2 Redshift survey (\cite{P02}) for the same magnitude selection but for a higher mean redshift : $f_p (-21 \leq M_B - 5\ log\ h\leq -18) = 3.21 \pm 0.77\%$ at $z = 0.3$.

de Propris et al (2007) derived measurements of the pair fraction using galaxy asymmetry and pair proximity to measure galaxy merger fractions for a volume limited sample of 3184 galaxies with $-21 \leq M_B - 5\ log\ h \leq -18$ and $0.010 \leq z \leq 0.123$ drawn from the Millennium Galaxy Catalogue.
They found a pair fraction of $4.1 \pm 0.4\%$ for galaxies with $r_p \leq 20h^{-1}\ kpc$.
 
Combining these values with our brighter sample ($M_B ({z=0}) \leq -18 +5 log(h) \sim -18.77$), we estimate $m = 1.50 \pm 0.76$ and $f_p(0) = 3.01 \pm 0.52$.
Here, we show that the fainter the galaxy sample is, the faster is the evolution of the pair fraction.
\begin{figure}[h!]
 \begin{center}
     \resizebox{!}{5.5cm}{\scalebox{1}{\input{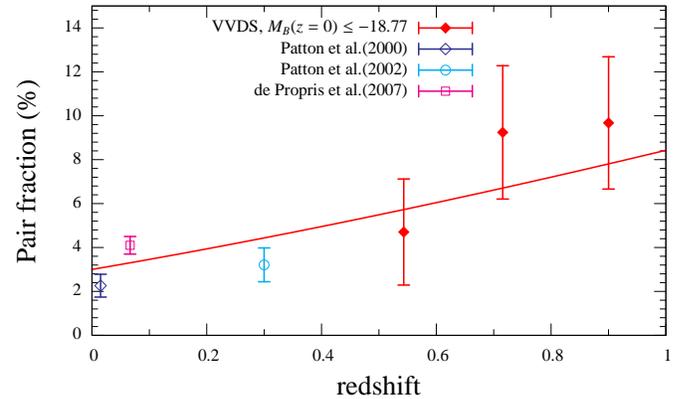}}}\\
    \caption{Evolution of the pair fraction as a function of redshift adding SSRS2 (open diamond), CNOC2 (open circle) and MGC (open square) low redshift points to VVDS measurements (filled diamonds) for $M_B(z=0) \leq -18.77$.}
    \label{low_redshift}
 \end{center}
    \end{figure} 

Figure~\ref{low_redshift} shows the best fit when combining these pair fraction measurements with our brightest sample.

 \subsection{Influence of stellar mass on the pair fraction}

To identify if the evolution of the pair fraction is also dependent on
the stellar mass of the galaxies (as a proxy for total mass), we applied exactly the same method as we used for the luminosity in Section \ref{3.1} but on a mass selected sample instead.
Using masses derived in the VVDS and the evolution of the characteristic stellar mass, $M^*_{star}$, as described in Pozzetti et al. (2007), we define a stellar-mass selected sample volume complete up to redshift $\sim 1$ (using an evolution parameter $Q_{Mass}(z)=-0.187 \times z$ to reproduce the evolution of $M^*_{star}$). Stellar masses are derived using a Bruzual \& Charlot (2003) model and allowing bursts on the top of a smooth star formation history. We applied the same corrections described in Section 2.4 by replacing the luminosity function by the mass function. 
We define a major pair via the ratio $M_1/M_2$ of stellar masses, and select pairs with $M_1/M_2 \leq 4$ corresponding roughly to a luminosity selected sample with $\Delta M_B \leq 1.5$ mag.

\begin{figure}[h!]
 \begin{center}
     \resizebox{!}{5.5cm}{\scalebox{1}{\input{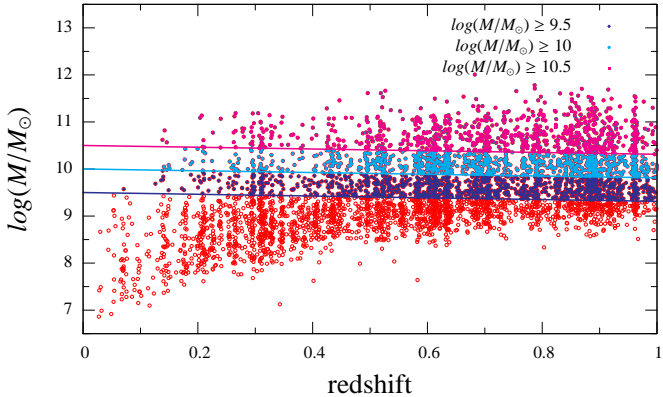}}}\\
    \caption{The three sub-samples defined to study the influence of the mass on the pair fraction using pairs with $r_p^{max} = 100h^{-1}\ kpc$.}
    \label{distrib_mass}
 \end{center}
    \end{figure} 

\begin{figure}[h!]
  \begin{center}
      \resizebox{!}{5.5cm}{\scalebox{1}{\input{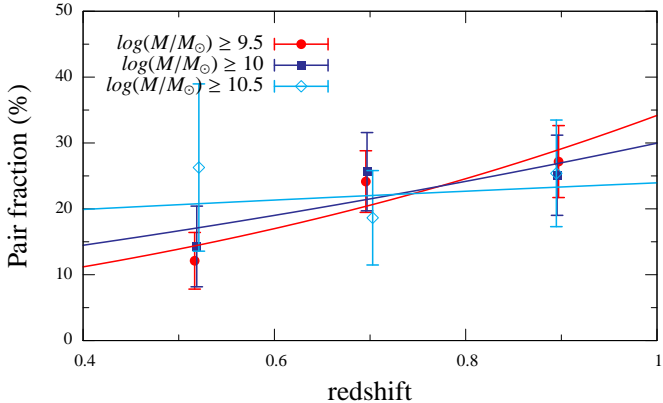}}}\\
     \caption{Evolution of the pair fraction for different sub-samples with different stellar mass limits
    $log(M/M_{\odot{}}) \geq 9.5$ (circles), $log(M/M_{\odot{}}) \geq 10$ (squares) and $log(M/M_{\odot{}}) \geq 10.5$ (empty diamonds).}
     \label{mass-pf}
  \end{center}
     \end{figure}

We divided our sample in different sub-samples: one with $log(M/M_{\odot{}}) \geq 9.5$ (106 pairs), one with $log(M/M_{\odot{}}) \geq 10$ (77 pairs) and one with $log(M/M_{\odot{}}) \geq 10.5$ (37 pairs) with separations $\Delta v \leq 500\ km/s$ and increasing the projected separation to $r_p^{max} = 100 h^{-1}\ kpc$ for better statistics, as shown in Figure~\ref{distrib_mass}.
Figure~\ref{mass-pf} shows the evolution of the pair fraction  in those different mass sub-samples. For low mass galaxies with $log(M/M_{\odot{}}) \geq 9.5$, $m = 3.13 \pm 1.54$ and $f_p(z=0) = 3.90 \pm 3.42$. For intermediate mass galaxies with $log(M/M_{\odot{}}) \geq 10$, $m = 2.04 \pm 1.65$ and $f_p(z=0) = 7.28 \pm 6.81$. For massive galaxies with $log(M/M_{\odot{}}) \geq 10.5$, $m = 0.52 \pm 2.07$ and $f_p(z=0) = 16.7 \pm 19.5$. We see a flatter evolution as we select more massive galaxies.
It is therefore apparent that intermediate or low mass galaxies are responsible for most of the evolution of the pair fraction and merger rate. 

  \begin{table*}[ht]
 \center
 \caption{Spectral types of pairs for the less and the most massive selected samples. Fractions are given in brackets.}
 \label{table_mass}
 \centering
 \begin{tabular}{c c c c c}

  &$N_{pairs}$&Early-type pairs&Late-type pairs&Mixed type pairs\\ \hline\hline \
$log(M/M_{\odot{}}) \geq 9.5$&$106$&$31\ (29.2\%)$&$53\ (50.0\%)$&$22\ (20.8\%)$\\  
$log(M/M_{\odot{}}) \geq 10$&$77$&$29\ (37.7\%)$&$29\ (37.7\%)$&$19\ (24.6\%)$\\
 $log(M/M_{\odot{}}) \geq 10.5$&$37$&$22\ (59.5\%)$&$6\ (16.2\%)$&$9\ (24.3\%)$\\
\end{tabular} 
\end{table*}

\section{Physical properties of galaxy pairs}

   \subsection{Spectro-photometric properties}

\begin{figure}[h!]
 \begin{center}
     \resizebox{!}{5.5cm}{\scalebox{1}{\input{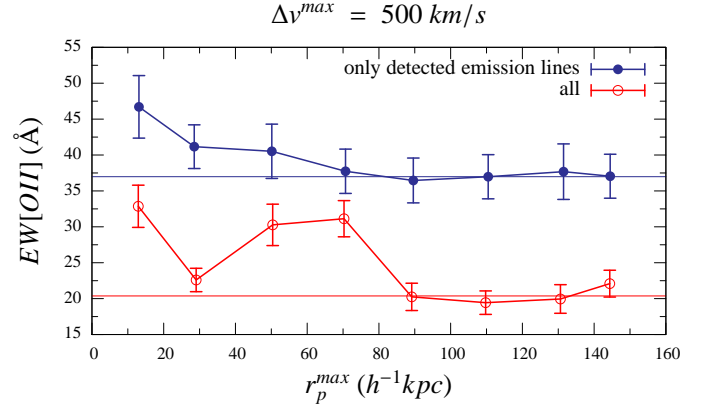}}}\\
    \caption{Mean $EW[OII]$ for pairs with $\Delta v ^{max}= 500\ km/s$ as a function of $r_p^{max}$. We present results using only galaxies where an $[OII]$ line has been detected (filled circles) and using all galaxies, including those where the line has not been detected (empty circles). Thin lines represent the mean value of $EW[OII]$ for the two sub-samples fitted on $80 <\ r_p^{max} <\ 150 h^{-1}kpc$.}
    \label{EW}
 \end{center}
    \end{figure}

  One of the expected effect of a merging or close interaction of galaxies is an increase 
in the star formation rate of the system. We evaluate here if our sample of
pairs has a stronger star formation rate than the global population by studying the rest-frame $[OII]3727{\AA}$ 
equivalent widths (EW) as a function of projected separation within a given $\Delta v^{max}$ ($500\ km/s$). 
$EW[0II]$ were derived using the $platefit$ 
software (\cite{lamareille}), which performs a continuum fit to the observed spectra using template fitting. 
It enables an unbiased measurement of the intensities of absorption and emission
lines. For each pair, we produced the mean $EW[OII]$ by summing the individual $EW[OII]$
of each galaxy, assuming $EW[OII]=0$ if the line is not detected. Using only galaxies for which the $[OII]$ line has been detected, the mean $EW[OII]$ is larger, on average, for the closest pairs with $EW[OII]=46.7 \pm 4.35$ for $r_p^{max} \leq 20h^{-1}\ kpc$, $EW[OII]=40.5 \pm 3.78$ for $r_p^{max} \leq 50h^{-1}\ kpc$, and $EW[OII]=36.5 \pm 3.12$ for $r_p^{max} \leq 100h^{-1}\ kpc$ indicating a $25.9 \pm 4.10\%$ increase in $EW[OII]$ at small separations (see Figure~\ref{EW}). We perform the same estimation using also galaxies for which the $[OII]$ line has not been detected ($EW[OII]=0$). Both samples show an increase of the mean $EW[OII]$ at small projected separations, extending to higher redshifts results of Woods et al. (2006) in the local CFA2 sample.
We conclude that star formation is enhanced in close merging systems at the mean redshift $<z>=0.76$ of our sample.

   \subsection{Spectral types of galaxies in pairs}

In this Section, we compare the spectral properties of galaxies in dynamical pairs with 
field galaxies. For each galaxy in the VVDS, the spectral type has been derived on the
basis of the template fitting of the rest-frame multi-$\lambda$ photometry (\cite{zucca}). Galaxies were classified in type 1 (E/S0), type 2 (Early spiral), type 3 (Late  spiral) and type 4 (Irregular). Therefore, for each pair, we know the
spectral types of both the primary galaxy and its companion(s).
 
We have investigated which galaxy types are involved in a pair as cosmic time 
evolves. We classified each pair with a flag $(X-X)$ where $X$ is the 
spectral type of each pair member. For instance 'dry mergers' with the 
merging of two early-type galaxies are classified as type $(1-1)$. We consider 
all the permutations between these four types. 
We classify as 'early-type' pairs, pairs with flags $(1-1)$, $(1-2)$ and 
$(2-2)$, late-type pairs the pairs with flags $(3-3)$, $(3-4)$, and
$(4-4)$, and mixed type pairs those with flags $(1-3)$, $(1-4)$, $(2-3)$,
and $(2-4)$. Table~\ref{class} gives the fraction 
of these different classes in the $r_p^{max}=100h^{-1}\ kpc$ pair sample.
The fraction of pairs involving only E/SO galaxies increases from $3.0\%$ at $z\sim 0.9$ to $11.8\%$ at $z\sim 0.5$, the fraction of pairs involving at least one E/SO increases from $22.4\%$ at $z\sim 0.9$ to $29.4\%$ at $z\sim 0.5$, while the vast majority of pairs involving at least one late-spiral or Irr galaxy represents a fraction decreasing from $83.4\%$ at $z\sim 0.9$ to $76.5\%$ at $z\sim 0.5$.

\begin{table*}[ht]
\center
\caption{Fraction of pairs vs. the spectral classes of each galaxy
in the pair for redshift $z \sim 0.5$ and $z \sim 0.9$. Pairs with $M_B < -18 - Q(z)$ and $r_p^{max}=100kpc/h$ 
are considered (202 pairs in total).}
\label{class}
\centering
\begin{tabular}{c c c c}

  & $Classification$  & $Fraction\ at\ z\sim 0.5$ & $Fraction\ at\ z\sim 0.9$\\ \hline\hline \
2 E/SO & $(1-1)$ & $11.8\%$ & $3.0\%$\\  
1 E/SO involved & $(1-X)$ & $29.4\%$ & $22.4\%$\\ 
1 E/SO or 1 early-Sp involved & $(1-X)+(2-X)$ & $47.1\%$ & $46.3\%$\\ \hline\hline \
2 Irr & $(4-4)$ & $17.6\%$ & $16.4\%$\\
1 Irr involved & $(4-X)$ & $47.1\%$ & $47.8\%$\\
1 Irr or 1 late-Sp involved & $(3-X)+(4-X)$ & $76.5\%$ & $83.4\%$\\
\end{tabular} 
\end{table*}

\begin{figure}[h!]
 \begin{center}
     \resizebox{!}{5.5cm}{\scalebox{1}{\input{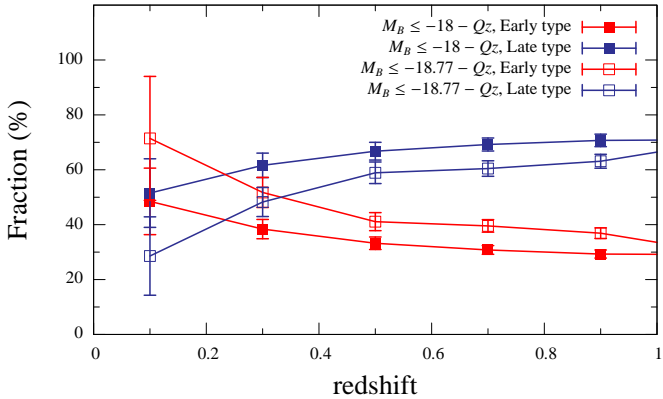}}}\\
    \caption{Fraction of early-type and late-type galaxies in the underlying selected sample brighter than 
$M_B = -18 - Q(z)$ (filled squares) and in the underlying selected sample brighter than
 $M_B = -18.77 - Q(z)$ (empty squares) as a function of redshift.}
    \label{spectype}
 \end{center}
    \end{figure}

\begin{figure}[h!]
 \begin{center}
     \resizebox{!}{5.5cm}{\scalebox{1}{\input{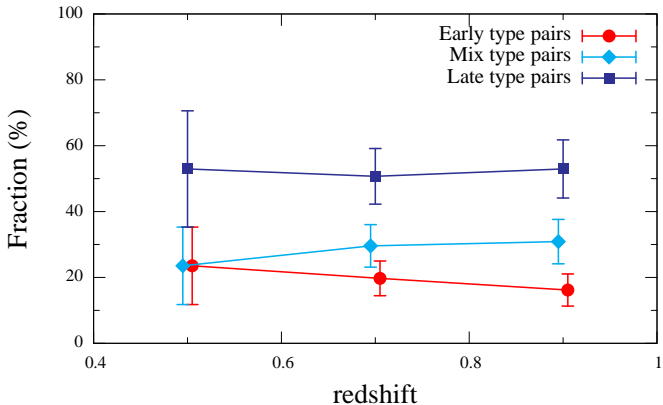}}}\\
    \caption{Fraction of early- (circles), mixed- (diamonds) and late-type (squares) pairs in the selected sample brighter than 
$M_B = -18 - Q(z)$ as a function of redshift.}
    \label{spectype-pairs}
 \end{center}
    \end{figure}

Figure~\ref{spectype} shows the evolution of the fraction of early-type, and late-type 
galaxies in two different samples : one brighter than $M_B = -18 - Q(z)$ 
(faint sample) and one brighter than $M_B = -18.77 - Q(z)$ (bright sample). 
In the faint sample, the population is dominated by late-type galaxies at all redshifts. 
In the bright sample, late-type galaxies dominate between $z\sim 0.4$ and $z\sim 1$, and early-type galaxies
become dominant between $z\sim 0.1$ and $z\sim 0.3$. 
Early-type galaxies represent only one third of the sample at $z\sim1$, but about two third at $z\sim0.1$. Figure~\ref{spectype-pairs} shows the fraction of early, mixed and late-type pairs with $M_B \leq -18 - Q(z)$ as a function of redshift. At $z \sim 0.9$, $15 \%$ ($55\%$) of these pairs are early(late) type pairs whereas at $z \sim 0.5$, $25 \%$ ($50\%$) of these pairs are early(late) type pairs following the same trend as the underlying sample of galaxies.

\begin{figure}[h!]
 \begin{center}
     \resizebox{!}{5.5cm}{\scalebox{1}{\input{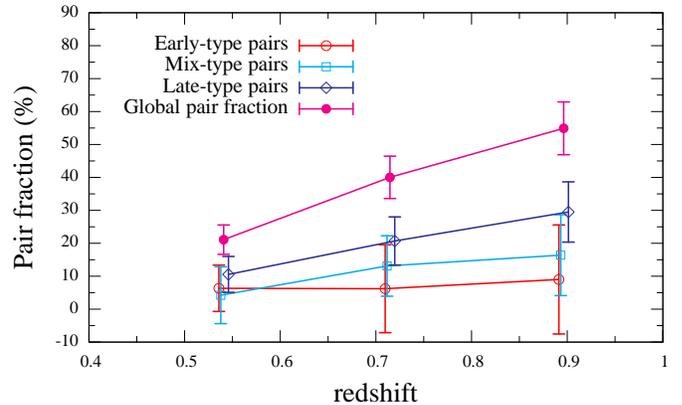}}}\\
    \caption{Evolution of the global pair fraction in the faint sample ($M_B(z=0) \leq -18$) with $r_p^{max} = 100h^{-1}\ kpc$ (pink) 
and contribution of early-types (red), mixed types (cyan) and late-type (blue).}
    \label{global-pair_fraction}
 \end{center}
    \end{figure}

Figure~\ref{global-pair_fraction} shows the pair fraction as a function of redshift, for
3 classes of pairs type compared to the global pair fraction: two early-type galaxies, one early and one late component, two late-types, using a sample with $r_p^{max} = 100h^{-1}\ kpc$. The early-type pair fraction evolves slowly with redshift with $m=1.44 \pm 0.93$. On the contrary, the late-early and late-type pair fractions evolve strongly with redshift, with $m=5.16 \pm 2.56$ and $m=4.74 \pm 0.81$ respectively.

Table~\ref{table_mass} gives the distributions of pairs as a function of the stellar mass selection and spectral types of the pairs. The $log(M/M_{\odot{}}) \geq 9.5$ sub-sample is dominated by late-type pairs (50\%) while the $log(M/M_{\odot{}}) \geq 10.5$ sub-sample is dominated by early-type pairs (59.5\%) (see Section 4.2).
We conclude that most of the pair fraction evolution is coming from lower mass late-type or mixed-type pairs.

\section{Evolution of the merger rate}

\begin{figure}[h!]
 \begin{center}
     \resizebox{!}{5.5cm}{\scalebox{1}{\input{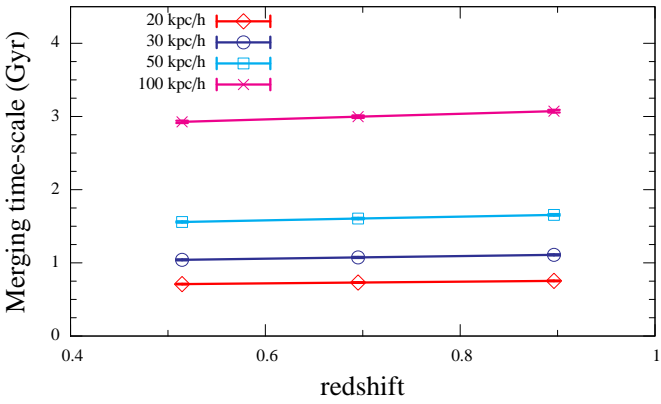}}}\\
    \caption{Evolution with redshift of the merging time-scale in Gyr as a function of $r_p^{max}$.}
    \label{timescale}
 \end{center}
    \end{figure}

 Knowing the pair fraction, we derive the merger rate i.e. the number of mergers per unit time and per comoving volume. This rate can be expressed as 
\begin{eqnarray}
N_{mg}(z) = C_{mg} \times \frac{\left(N_{p}^{corr}-N_{triplets}^{corr}\right)}{N_{g}^{corr}} \times n(z) \times T_{mg}^{-1} 
\end{eqnarray}
where $C_{mg}$ stands for the fraction of galaxies in close pairs that will undergo a merger within the time $T_{mg}$ and $n(z)$ is the comoving number density of galaxies. 
The best way to estimate these values is to use simulations to follow the merging history of galaxies with different masses. 
We take results from the Millennium simulations  (Kitzbichler \& White, 2008) to estimate the merging time-scale $T_{mg}(r_p^{max},z)$, written as follows: $$T_{mg}^{-1/2} = T_0(r_p^{max})^{-1/2} + f_1(r_p^{max}) \times z + f_2(r_p^{max}) \times (logM_{*}-10).$$
We computed $T_0, f_1$ and $f_2$ for $r_p^{max} = (20,\ 30,\ 50\ and\ 100)h^{-1}\ kpc$ in the case of $\Delta v^{max} = 500\ km/s$. 
Following Lin et al. (2008), the probability for a pair to merge in the given time-scale $T_{mg}$ is assumed constant, $C_{mg} = 0.6$, independent of the separation $r_p^{max}$. As a proxy for total mass, we use the evolution of the characteristic stellar mass $M_{stars}^{*}$ as derived in the VVDS (\cite{pozzetti}). Figure~\ref{timescale} shows the change in the galaxy merging time-scale with redshift and $r_p^{max}$. 

The time-scales are found to be higher than the standard assumption that half of the pairs with $r_p^{max} = 20h^{-1}\ kpc$ undergo a merger in half a Giga-year (\cite{P00}; \cite{P02}; \cite{lin04}). Using the Kitzblicher and White (2008) prescription, we find that even for the closest pairs the merging time-scale is 1.5 times higher than assumptions previously used in the literature.

\begin{table*}[htbp]
\caption{Merger rate values for different sets of parameter and redshift in units of $10^{-4}\ $mergers $Mpc^{-3} Gyr^{-1}$ for the bright sample ($M_B^{max}(z=0) = -18$) for galaxies with $\Delta M_{B} \leq 1.5$.}
\label{values_merger}
\centering
\begin{tabular}{|c c c c c|}
\hline
\multicolumn{5}{|c|}{$\mathbf{\Delta v \leq 500\ km/s}$}\\ \hline
 & $20 h^{-1} kpc$ & $30 h^{-1} kpc$ & $50 h^{-1} kpc$ & $100 h^{-1} kpc$ \\ \hline 
$z_{mean}=0.5124$ & $8.17 \pm 3.87$ & $8.95 \pm 3.39$ & $14.43 \pm 3.56$ & $12.32 \pm 2.41$ \\ 
$z_{mean}=0.6952$ & $15.60 \pm 5.97$ & $14.10 \pm 4.73$ & $14.18 \pm 3.85$ & $17.91 \pm 3.29$ \\ 
$z_{mean}=0.8989$ & $14.08 \pm 6.47$ & $13.48 \pm 5.27$ & $19.31 \pm 5.21$ & $19.37 \pm 3.90$ \\ 
\hline
\multicolumn{5}{|c|}{$\mathbf{\Delta v \leq 1000\ km/s}$}\\ \hline
 & $20 h^{-1} kpc$ & $30 h^{-1} kpc$ & $50 h^{-1} kpc$ & $100 h^{-1} kpc$ \\ \hline
$z_{mean}=0.5124$ & $13.94 \pm 5.15$ & $12.89 \pm 4.13$ & $17.92 \pm 4.01$ & $18.69 \pm 3.07$ \\ 
$z_{mean}=0.6952$ & $20.59 \pm 6.95$ & $18.32 \pm 5.47$ & $18.93 \pm 4.53$ & $23.02 \pm 3.81$ \\ 
$z_{mean}=0.8989$ & $17.44 \pm 7.26$ & $17.35 \pm 6.05$ & $24.27 \pm 5.91$ & $23.51 \pm 4.37$ \\ 
\hline
\multicolumn{5}{|c|}{$\mathbf{\Delta v \leq 2000\ km/s}$}\\ \hline
 & $20 h^{-1} kpc$ & $30 h^{-1} kpc$ & $50 h^{-1} kpc$ & $100 h^{-1} kpc$ \\ \hline
$z_{mean}=0.5124$ & $15.67 \pm 5.49$ & $14.07 \pm 4.33$ & $22.58 \pm 4.55$ & $27.17 \pm 3.78$ \\ 
$z_{mean}=0.6952$ & $20.59 \pm 6.95$ & $20.14 \pm 5.77$ & $21.03 \pm 4.78$ & $26.48 \pm 4.10$ \\ 
$z_{mean}=0.8989$ & $18.65 \pm 7.52$ & $18.17 \pm 6.20$ & $25.53 \pm 6.05$ & $27.21 \pm 4.70$ \\
\hline
\end{tabular}
\end{table*}

\begin{figure}[h!]
 \begin{center}
     \resizebox{!}{5.5cm}{\scalebox{1}{\input{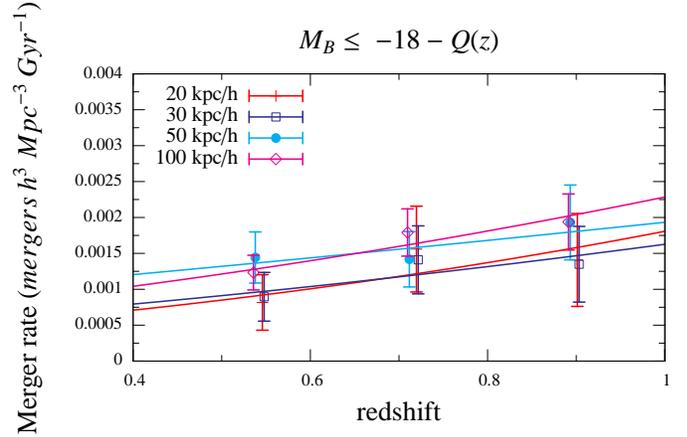}}}\\
    \caption{Evolution of the $M_B < -18 - Q(z)$ galaxy merger rate for different sets of $r_p^{max}$ with $\Delta v ^{max}= 500 km/s$.}
    \label{fit-merger_rate}
 \end{center}
    \end{figure}

The merger rate should be an ''absolute value'', independent of $r_p^{max}$ and $\Delta v$ since we take into account the merging time-scales corresponding to different pair separations. To check that the merger rate does not depend on the adopted value of $r_p^{max}$ and $\Delta v$, we have computed the merger rate for different sets of $r_p^{max}$ with $\Delta v \leq 500 km/s$; results are presented in Table~\ref{values_merger}, and plotted in Figure~\ref{fit-merger_rate}. 
The merger rate values are in good agreement, both in slope and normalization, for different sets of projected separations. 
This is a good indication of the robustness of the method. In the following, we use values of the merger rate with $r_p^{max} = 100h^{-1} kpc$ for better statistics, when necessary.
The merger rate increases from $\sim 12.3 \times 10^{-4}$ to $\sim 19.4 \times 10^{-4}\ $ mergers $\ h^{3}\ Mpc^{-3}\ Gyr^{-1}$ from $z=0.5$ to $z=0.9$.
The merger rate evolves as $N_{mg}(z) = N_{mg}(z=0) \times (1+z)^{m_{mg}}$ with $m_{mg} = 2.20 \pm 0.77$ and $N_{mg}(z=0) = (4.96 \pm 2.07) \times 10^{-4}$.
Table~\ref{evol_merger} lists the values of the parameters $m_{mg}$ and $N_{mg}(z=0)$ for different sets of separations and we plot in Figure~\ref{fit-merger_rate} the evolution of the merger rate for $r_p^{max} = 20,\ 30,\ 50,\ 100h^{-1}\ kpc$ and $\Delta v^{max}=500 km/s$.

Using this merger rate evolution parametrisation, we estimate the 
fraction of present day galaxies $f_{rem}$, that have undergone one major merger (\cite{P02}) since $z \sim 1$.
\begin{eqnarray}
f_{rem}=1-\prod_{j=1}^{N} \frac{1-f_{mg}(z_j)}{1-0.5 f_{mg}(z_j)}
\end{eqnarray}
where $z_j$ corresponds to a lookback time of $t=j\times T_{mg}$ and $f_{mg}$ is 
the fraction of galaxies that undergo a merger. We use the merger rate evolution 
derived with $r_p^{max}=20h^{-1}\ kpc$ and lookback times derived using 
\begin{eqnarray}
t_{lookback}=\frac{c}{H_0}\int_{0}^{z}\frac{dz'}{(1+z')\sqrt{\Omega_m(1+z')^3 + \Omega_{\Lambda}}}
\end{eqnarray}
based on a mean merging time-scale of 0.75 Gyr (corresponding to $r_p^{max}=20h^{-1}\ kpc$). We find that $8\%$ of present day galaxies brighter than $M_B = -18 - Q(z)$ have undergone a major merger since $z \sim 0.4$, , while $22\%$ have done so
since $z \sim 0.9$.

\begin{table*}[htbp]
\caption{Best fit parameters of $m_{mg}$ and $N_{mg} (z=0)$ for major mergers as a function of the dynamical parameters for the faint sample and for galaxies with $\Delta M_{B} \leq 1.5$. .}
\label{evol_merger}
\centering
\begin{tabular}{|c c c c c|}
\hline
\multicolumn{5}{|c|}{$\mathbf{\Delta v \leq 500\ km/s}$}\\ \hline
 $r_p$ & $20 h^{-1} kpc$ & $30 h^{-1} kpc$ & $50 h^{-1} kpc$ & $100 h^{-1} kpc$ \\ 
$m_{mg}$ & $2.63 \pm 1.96$ & $2.01 \pm 1.32$ & $1.33 \pm 0.98$ & $2.20 \pm 0.77$ \\ 
$N_{mg} (z=0) \times 10^{-4}$ & $2.93 \pm 3.14$ & $4.03 \pm 2.88$ & $7.70 \pm 4.05$ & $4.96 \pm 2.07$ \\ 
\hline
\multicolumn{5}{|c|}{$\mathbf{\Delta v \leq 1000\ km/s}$}\\ \hline
 $r_p$ & $20 h^{-1} kpc$ & $30 h^{-1} kpc$ & $50 h^{-1} kpc$ & $100 h^{-1} kpc$ \\ \hline
$m_{mg}$  & $1.19 \pm 1.48$ & $1.50 \pm 1.08$ & $1.42 \pm 0.58$ & $1.15 \pm 0.50$\\ 
$N_{mg} (z=0) \times 10^{-4}$ & $8.98 \pm 7.12$ & $7.14 \pm 4.16$ & $9.40 \pm 2.94$ & $11.7 \pm 3.10$ \\ 
\hline
\multicolumn{5}{|c|}{$\mathbf{\Delta v \leq 2000\ km/s}$}\\ \hline
 $r_p$ & $20 h^{-1} kpc$ & $30 h^{-1} kpc$ & $50 h^{-1} kpc$ & $100 h^{-1} kpc$ \\ \hline
$m_{mg}$ & $0.93 \pm 1.00$ & $1.31 \pm 1.23$ & $0.49 \pm 0.78$ & $-0.02 \pm 0.15$ \\ 
$N_{mg} (z=0) \times 10^{-4}$ & $11.0 \pm 5.89$ & $8.49 \pm 5.61$ & $17.5 \pm 7.24$ & $27.2 \pm 2.13$ \\ 
\hline
\end{tabular} 
\end{table*}

We have also computed the merger rate for two different luminosities using pairs with $r_p^{max} = 100h^{-1} kpc$.
A similar trend to the pair fraction is observed: for galaxies with $M_B (z=0) \leq -18$,\ we find $m_{mg} = 2.20 \pm 0.77$, while for brighter galaxies with  $M_B (z=0) \leq -18.77$ we find $m_{mg} = 1.60 \pm 1.83$ using only VVDS data. For the same limiting magnitude and using the merger rate measured by de Propris et al. (2007) to constrain the low redshift end, $m_{mg} = 1.57 \pm 0.44$.

\begin{figure}[htbp!]
  \begin{center}
        \resizebox{!}{5.5cm}{\scalebox{1}{\input{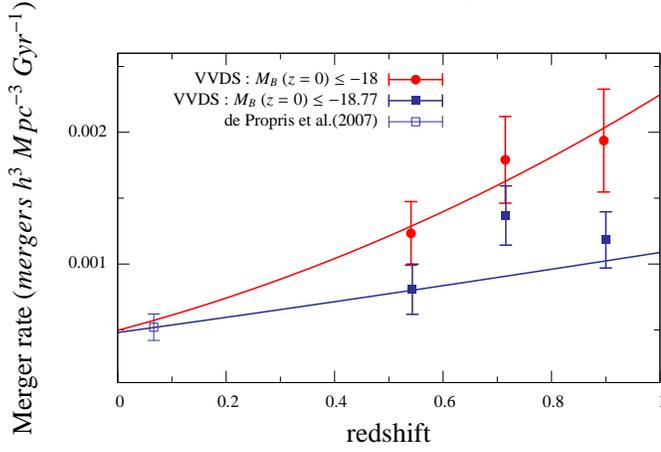}}}\\
      \caption{Evolution of the merger rate for different luminosity ranges. For the brightest sample ($M_B(z=0)\ \leq\ -18.77$), we add results from de Propris et al. (2007) (empty square)}
      \label{merger_rate_MB}
 \end{center}
  \end{figure}

\begin{figure}[htbp!]
  \begin{center}
        \resizebox{!}{5.5cm}{\scalebox{1}{\input{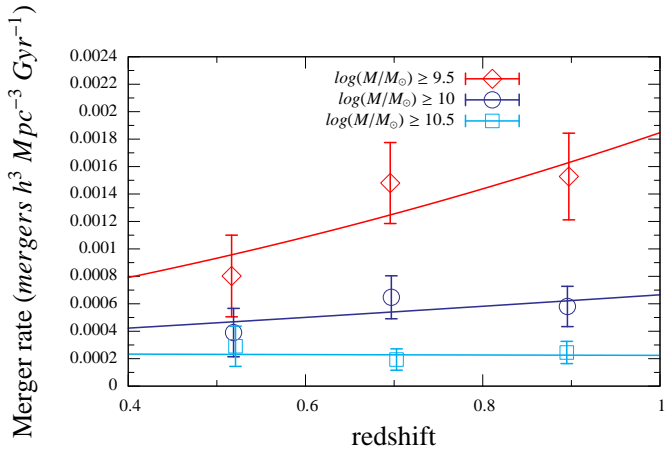}}}\\
 \caption{Evolution of the merger rate for different mass ranges using $r_p^{max} = 100h^{-1} kpc$. From top to bottom : $log(M/M_{\odot{}}) \geq 9.5$ (diamonds), $log(M/M_{\odot{}}) \geq 10$ (circles) and $log(M/M_{\odot{}}) \geq 10.5$ (squares)}
      \label{merger_rate_mass}
 \end{center}
  \end{figure}

Similarly, we have computed the merger rate for different mass selected samples as defined in Section 3.3 using $r_p^{max} = 100 h^{-1}kpc$. For the less massive sample ($log(M/M_{\odot{}}) \geq 9.5$), $m_{mg} = 2.38 \pm 1.57$ with $N_{mg} (z=0) = (3.56 \pm 3.17) \times 10^{-4}\ $ mergers$\ h^3 Mpc^{-3} Gyr^{-1}$, while for the intermediate sample ($log(M/M_{\odot{}}) \geq 10$), $m_{mg} = 1.27 \pm 1.67$ with $N_{mg} (z=0) = (2.75 \pm 2.61) \times 10^{-4}\ $ mergers$\ h^3 Mpc^{-3} Gyr^{-1}$, as shown in Figure~\ref{merger_rate_mass}.

We see a change in the evolution of the merger rate as we go to the highest masses. First, the number of less massive merging events ($log(M/M_{\odot{}}) \geq 9.5$) is greater than the number of high mass merging events ($log(M/M_{\odot{}}) \geq 10.5$). Then we see a flattening of the evolution of the merger rate as we go to higher mass galaxies, confirming that the evolution of the major merger rate is mainly due to the less massive galaxy population.

\section{Stellar mass involved in mergers}

\begin{figure}[htbp!]
  \begin{center}
        \resizebox{!}{5.5cm}{\scalebox{1}{\input{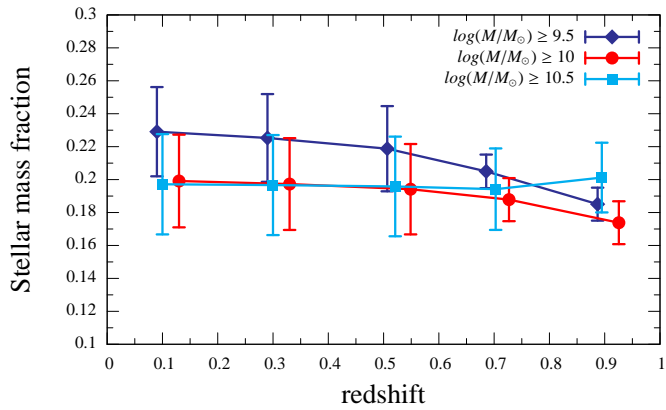}}}\\
   \caption{The fraction of stellar mass density involved in a merger process since $z \sim 1$ as a function of redshift for different mass selected sub-samples.}
      \label{fraction_stellar_mass}
 \end{center}
  \end{figure}

We estimate the fraction of the total stellar mass involved in a merger process, $f_{M^*}(z)$, since $z \sim 1$ as a function of redshift as
\begin{eqnarray}
f_{M^*} (z)= \frac{M_{merger}^{*}(z) \times N_{mg}(z) \times T_{bin}(z)}{M_{tot}^{*}(z) \times n(z)},
\end{eqnarray}
where $n(z)$ is the comoving number density of galaxies, $N_{mg}(z)$ is the number of mergers per unit of time and per comoving volume, $M_{merger}^{*}(z) = \frac{\sum M_1 + M_2}{N_{pairs}(z)}$ is the mean stellar mass involved in a merger process, $T_{bin}(z)$ is the elapsed time corresponding to the considered redshift bin and $M_{tot}^{*}(z)$ is the total stellar mass in the redshift interval.
To extrapolate the values of the stellar mass densities at $z \sim 0.1$, we assumed a constant stellar mass density below $z=0.4$. This assumption is consistent with the evolution of $\rho_{*}$ reported in Pozzetti et al. (2007).
We show in Figure~\ref{fraction_stellar_mass} that around $25\%$ of the stellar mass contained in galaxies with $log (M/M_{\odot_{}}) \geq 9.5$ at $z \sim 0.1$ have experienced a merger since $z \sim 1$ while this fraction is about $20\%$ for galaxies with $log (M/M_{\odot_{}}) \geq 10$. One can identify two trends: the fraction of the stellar mass density coming from the merging process shows a rise of about $24\%$ from $z \sim 0.9$ down to $z \sim 0.1$ for the less massive population,
 whereas it stays roughly constant at about $20\%$ for the most massive galaxies.

\section{Summary and Discussion}

\label{discuss}

%
%

Our results can be summarized as follows: 

(i) We find that $3.8 \pm 1.7$, $9.4 \pm 2.8$, and $10.9 \pm 3.2\ \%$ of galaxies with $M_B(z) < -18 - Q(z)$ at $z \sim 0.5, 0.7$ and $0.9$ respectively, are in pairs of galaxies with luminosities $\Delta M_B \leq 1.5$ and separations less than $20h^{-1}\ kpc$. 

(ii) The evolution of the pair fraction with redshift is strongly dependent on the absolute luminosity or stellar mass of the brighter galaxy in the pair: it evolves more slowly for brighter or more massive galaxies than for faint galaxies. Using the VVDS alone,
the pair fraction of galaxies with  $M_B(z) < -18 -Q(z)$ is found to strongly evolves with
redshift as $\propto (1+z)^m$ with $m=4.46 \pm 0.81$ for separations of $(100h^{-1}\ kpc,500 km/s)$, while for brighter galaxies with $M_B(z) < -18.77 -Q(z)$, we find a slower evolution with $m=3.18 \pm 1.34$. Combining VVDS data with low redshift measurements from de Propris et al. (2007), Patton et al. (2000) and Patton et al. (2002), and taking $r_p^{max}=\ 20h^{-1}\ kpc$, we similarly find $m=1.50 \pm 0.76$ for bright galaxies with $M_B(z=0) \leq -18 + 5log(h) \sim -18.77$ and $m=4.73 \pm 2.01$ for the fainter $M_B(z=0) \leq -18$ sample. In addition, the evolution of the pair fraction is found to be stronger with $m = 3.13 \pm 1.54$ for less massive galaxies with $log(M/M_{\odot}) \geq 9.5$, than for more massive galaxies with $log(M/M_{\odot}) \geq 10$ for which we find $m = 2.04 \pm 1.65$. Low mass pairs are therefore contributing more to the evolution of the pair fraction than high mass pairs.

(iii) The star formation rate of close pairs is enhanced at separations $r_p \leq 
150h^{-1}\ kpc$. We find that the mean $EW(OII)$ in close pairs are larger by $26 \pm 4\%$ than the one derived for galaxies with larger separations.

(iv) The evolution of the pair fraction is stronger for late-type pairs with $m_{late} = 4.74 \pm 0.81$, than for early-type pairs with $m_{early} = 1.44 \pm 0.93$. Late-type pairs are therefore contributing significantly more to the observed evolution of the pair fraction than early-type pairs in our $I_{AB} \leq 24$ sample.

(v) Using the merging timescale from Kitzbichler \& White (2008), we find that the merger rate increases from $\sim 12.3 \times 10^{-4}$ to $\sim 19.4\times 10^{-4}\ $mergers$\ h^{3}\ Mpc^{-3}\ Gyr^{-1}$ from $z=0.5$ to $z=0.9$. The merger rate of galaxies with $M_B(z) < -18 -Q(z)$ evolves as $N_{mg}=(4.96 \pm 2.07) \times 10^{-4}) \times (1+z)^{2.20 \pm 0.77}$. Similarly to the pair fraction, we find that the merger rate evolves faster for fainter or less massive galaxies, with $m_{mg}=2.20 \pm 0.77$ and $2.38 \pm 1.57$ respectively, than for brighter or more massive galaxies with $m_{mg}=1.57 \pm 0.44$ and $1.27 \pm 1.67$ respectively. The merger rate is evolving more strongly for late-type mergers than for early-type mergers.


We conclude that the observed evolution of the pair fraction and merger rate in our $I_{AB} \leq 24$ sample is mostly driven by low mass late-type galaxies, while the pair fraction and merger rate of high mass early-type galaxies remains roughly constant since $z \sim 1$. Therefore, the pair fraction or the merger rate are not universal numbers but rather are dependent on the luminosity or stellar mass, and on the spectral type of galaxies involved. Our finding that bright or massive galaxies experience a lower merger rate and a lower evolution of the merger rate extends to higher redshifts the results found in the local Universe by Patton \& Atfield (2008). Taking into account this pair fraction and merger rate dependancy on galaxy luminosity and spectral type offers a first step to reconcile apparently inconsistent observations.Lotz et al. (2008) find a slow or no evolution of the merger rate and claim that they disagree with previous studies. When taking into account that their result is derived from bright $M_B \leq -19.94 - 1.3 \times z$ galaxies, their result is consistent with other studies like Conselice et al. (2003) or Le F\`evre et al. (2000) which have been analysing fainter samples.

 The dependency of the merger rate and its evolution on luminosity or stellar mass is indeed a prediction from the latest simulations using advanced semi-analytic models as described in Kitzbichler and White (2008). At the limiting magnitudes or stellar masses of our sample, Kitzbichler and White (2008) predict that the merger rate decreases and evolves more slowly for galaxy samples with increasing luminosity or stellar mass, similar to the trend observed in our sample. 

The star formation rate is significantly enhanced in merging pairs with a net star formation increase of $\sim 25\%$ for these galaxies. Nevertheless, it accounts for only $12\%$ to $3\%$ of the global galaxy population from redshift $z \sim 1$ to $z \sim 0$ which is not sufficient to counteract the strong fading of the global star formation rate observed since $z\sim1$. This may indicate that the gas reservoir of massive and intermediate mass galaxies has already been depleted at redshifts $z \sim 1$, in agreement with
their observed peak in star formation at $z\sim 3.5$ (e.g. Tresse et al., 2007). It is then apparent that the decreasing SFR since
$z\sim 1$ is regulated by other physical processes like gas availability in the intergalactic medium, or feedback.

Major merging events are largely dominated by pairs of late or mixed type galaxies, but while early-type mergers represent about $15\%$ of the merging events of bright galaxies at $z \sim 1$, they become approximately $25\%$ of all mergers at $z \sim 0.5$, which is in good agreement with previous results on dry mergers (e.g. \cite{lin08}). This indicates that major mergers are efficient in lowering the number density of intermediate
mass late-type galaxies to build up more early-type galaxies. 
We confirm that merging is one of the important physical processes driving galaxy evolution, with
the observed galaxy merger rate undoubtedly closely linked to the hierarchical build up of dark matter galaxy halos,
with a rapid mass accretion phase of massive halos since $z\sim1$ (Abbas et al., 2008). Our finding that $\sim20\%$ of the
stellar mass in present day massive galaxies has experienced a major merger since $z \sim 1$ is an indication that major
mergers are significantly contributing to the observed evolution of the stellar mass density
since $z\sim1$ (Bundy et al., 2005; Arnouts et al., 2007).

\begin{acknowledgements}
We are thankful to Simon White for forwarding us a preprint of the Kitzbichler and White (2008) work ahead of publication.
This research has been developed within the framework of the VVDS consortium.\\
This work has been partially supported by the
CNRS-INSU and its Programme National de Cosmologie (France),
and by Italian Ministry (MIUR) grants
COFIN2000 (MM02037133) and COFIN2003 (num.2003020150) 
and by INAF grants (PRIN-INAF 2005).\\
The VLT-VIMOS observations have been carried out on guaranteed
time (GTO) allocated by the European Southern Observatory (ESO)
to the VIRMOS consortium, under a contractual agreement between the
Centre National de la Recherche Scientifique of France, heading
a consortium of French and Italian institutes, and ESO,
to design, manufacture and test the VIMOS instrument.

Based on observations obtained with MegaPrime/MegaCam, a joint  
project of CFHT and CEA/DAPNIA, at the Canada-France-Hawaii Telescope  
(CFHT) which is operated by the National Research Council (NRC) of  
Canada, the Institut National des Science de l'Univers of the Centre  
National de la Recherche Scientifique (CNRS) of France, and the  
University of Hawaii. This work is based in part on data products  
produced at TERAPIX and the Canadian Astronomy Data Centre as part of  
the Canada-France-Hawaii Telescope Legacy Survey, a collaborative  
project of NRC and CNRS.

\end{acknowledgements}


\end{document}